\documentclass[epj,nopacs]{svjour}
\usepackage{amsmath,amssymb,amsfonts,latexsym,dcolumn,graphics}
\begin{document}
\title{Non-equilibrium Lifshitz theory as a steady state of a full dynamical quantum system}
\author{Fernando C. Lombardo\inst{1} \and Francisco D. Mazzitelli\inst{2} \and Adri\'an E. Rubio L\'opez\inst{1} \and Gustavo J. Turiaci\inst{3}
}                     
%
%
\institute{Departamento de F\'\i sica {\it Juan Jos\'e
Giambiagi}, FCEyN UBA and IFIBA CONICET-UBA, Facultad de Ciencias Exactas y Naturales,
Ciudad Universitaria, Pabell\' on I, 1428 Buenos Aires, Argentina \and Centro At\'omico Bariloche and Instituto Balseiro,
Comisi\'on Nacional de Energ\'\i a At\'omica, R8402AGP Bariloche, Argentina \and Physics Department, Princeton University, Princeton, NJ 08544, USA \\
\email{arubio@df.uba.ar}}
\date{Received: date / Revised version: date}
%
\abstract{
In this work we analyze the validity of Lifshitz's theory for the case of non-equilibrium scenarios from a full quantum dynamical approach. We show that Lifshitz's framework for the study of the Casimir pressure is the result of considering the long-time regime (or steady state) of a well-defined fully quantized problem, subjected to initial conditions for the electromagnetic field interacting with real materials. For this, we implement the closed time path formalism developed in previous works to study the case of two half spaces (modeled as composite environments, consisting in quantum degrees of freedom plus thermal baths) interacting with the electromagnetic field. Starting from initial uncorrelated free subsystems, we solve the full time evolution, obtaining general expressions for the different  contributions to the pressure that take part on the transient stage. Using the analytic properties of the retarded Green functions, we obtain the long-time limit of these contributions to the total Casimir pressure. We show that, in the steady state, only the baths' contribute, in agreement with the results of previous works, where this was assumed without justification. We also study in detail the physics of the initial conditions' contribution and the concept of modified vacuum modes, giving insights about in which situations one would expect a non vanishing  contribution at the steady state of a non-equilibrium scenario. This would be the case when considering  finite width slabs instead of half-spaces.
} 
\maketitle
%
\section{Introduction}

In this paper we are concerned with a first principles calculation of the non-equilibrium Casimir pressure between real materials, for the particular geometry of half-spaces separated by a vacuum gap of constant width. By non-equilibrium we refer to a situation in which the objects are held at different (but fixed) temperatures.

Lifshitz formula describes the Casimir pressure in a steady situation, in terms of the macroscopic properties of the materials. The original derivation  \cite{Lifshitz} was based on the use of stochastic Maxwell equations, and relied heavily on the thermal properties of the stochastic sources, assuming thermal equilibrium. A few years later, the same formula was obtained in Ref.\cite{Lifshitz2} within the framework of quantum field theory at finite temperature. Whatever the chosen approach, the final result can be expressed in terms of the permittivity of the materials (or alternatively in terms of their reflection coefficients), and therefore admits a natural generalization to the case in which the half-spaces are maintained at different temperatures, assuming that the system reaches a steady state. This kind of approach has been followed in Ref.\cite{Antezza}, where the non-equilibrium Casimir force has been considered for the first time. Nevertheless, the latter seems rather an extension of the original derivation than a full quantum development from first principles which is, today, a pending challenge in non-equilibrium scenarios. Although there are remarkable works in the context of Casimir physics (as Ref.\cite{BehuninHu2011}), the present paper focuses on the dynamical aspects of the building-up of the steady state of an electromagnetic (EM) field interacting with matter in a non-equilibrium scenario that, as far as we know, has not been done before,  and we think that contains relevant physics.

Beyond the non-equilibrium features, this work also enters another (not completely unrelated) issue that has been the subject of an intense controversy: whether Lifshitz theory is applicable or not to real metals. In the simplest approach, free (conduction) electrons in metals are described by the Drude model, although its dissipationless limit, the plasma model, seems to be in concordance with the experimental results \cite{Mostep2015}. Some authors claim that Lifshitz formula cannot be applied for the Drude model \cite{Mostep2009}, while others have reached the opposite conclusion \cite{Guerout2014}, claiming that the Lifshitz-Matsubara formula is not correct for the plasma model. At first glance, our work will not address directly this important question, because thro- ughout the paper we consider a material without free carriers, i.e., an insulator which permittivity function is given by the bounded electrons only (see Ref.\cite{Mostep2009}, for example). However, the present work enters the discussion about a systematic treatment that allows to understand the limitations of the Lifshitz formula, and how to figure out which model is consistent with it.

A first principles calculation involves a microscopic model for the quantum polarization degrees of freedom of the materials, interacting both with the electromagnetic field and thermal baths that fix the different temperatures of the objects (see Ref.\cite{LombiMazziRL,KnollLeonhardt1992} for more details about physical aspects of the material model). The evolution of the fully quantized system from adequate initial conditions should eventually show the emergence of a steady situation in the long time limit, with Casimir pressure given by a Lifshitz-like formula. From a technical point of view, this program can be developed in the framework of the theory of quantum open systems (see Ref.\cite{Petruccione}). The quantization procedure can be worked out through Heisenberg equations for the quantum operators (canonical formalism \cite{Greiner}) or path integral methods (closed time path formalism (CTP) \cite{CalHu}), but both are subject to initial conditions \cite{Milonni}. For the latter, after integration of the material degrees of freedom and thermal baths, one obtains the influence action (Refs.\cite{CalHu,FeynmanHibbs}) for the electromagnetic field. Then, integrating the field, one can construct the generating functional to calculate the correlation functions (see Ref.\cite{FradkinGitman}). Using the correlation functions derived, one can compute the time evolution of the mean value of the energy momentum tensor, and therefore the Casimir pressure as a function of time. The main question is therefore whether this mean value reaches a well defined limit at long times, and whether this limit depends on the initial conditions or not. Note that  the approaches that are usually  implemented in Casimir physics can be considered as `steady' quantization schemes in the sense that they assume without justification that the system reaches a steady state and, in addition, that the latter is of thermal equilibrium. Both requirements makes the `steadiness' assumption physically reasonable and expected to treat equilibrium situations (see for example Refs.\cite{Milonni,EberleinRobaschik2006} for the `steady canonical quantization' schemes and Ref.\cite{Bechler} for the `steady path integral quantization'). Nonetheless, it is not clear how to extend it to non-equilibrium (steady) situations or how it is done from first principles. As we mentioned before, the first work in this direction was Ref.\cite{Antezza} which extends the stochastic approach, but a complete quantum approach will bring to light the physics of the contributions of each part of the model to the electrodynamical quantities. At the end, the whole picture gives an appropriate and systematic way to deal with non-equilibrium situations in Casimir physics.

In previous works, some of us investigated preliminary aspects of the problem at hand. In Ref.\cite{CTPScalar} the case of  a quantum scalar field
in the presence of an arbitrary material was studied. After developing the quantum open systems approach for the particular situation of a scalar field interacting with microscopic degrees of freedom and thermal baths, it was shown that the emergence of a steady state independent of the initial conditions is a nontrivial issue. While for a quantum field in bulk material such limit exists, this is not the case for slabs of material with finite width. The generalization to the electromagnetic field has been considered in Ref.\cite{CTPGauge}, where the CTP generating functional for the electromagnetic field interacting with a composite environment has been described in detail. Specifically, the environment considered there consisted of the quantum polarization degrees of freedom at each point of space, connected to thermal baths to fix the local temperature. Formal expressions for the Maxwell tensor and the Poynting vector in terms of the Hadamard propagator were obtained, for the case of the electromagnetic field in bulk material, where, once again, the steady state is independent of the initial conditions. Some technical issues regarding the gauge fixing procedure in the context of CTP have also been discussed.

In the present paper we will extend the results in \cite{CTPGauge} to the case of the electromagnetic field in the presence of two half-spaces of real media separated by a vacuum gap. Considering the previous results mentioned above, it is not clear a priori wether the initial conditions will be erased or not in the long time limit for this particular geometry. Using the formalism of Ref.\cite{CTPGauge} we will obtain explicit expressions for the Casimir pressure as a function of time, assuming that the interaction of the vacuum field with the materials starts at a given time $t_0$. The Casimir pressure will receive three different contributions, coming from the field's initial conditions, the polarization degrees of freedom, and from the thermal baths. We will show that, in the long time limit, only the baths' contribution survive. This is the contribution expected in the Lifshitz approach. Therefore, we will provide a first principles derivation of Lifshitz formula for non-equilibrium situations in the framework of quantum open systems.

The paper is organized as follows: In the next Section we present the model and the influence action that includes noise and dissipation contributions to the calculation of the Casimir force. Section III contains the expression of the Casimir pressure in terms of the electromagnetic correlation functions and, following this, the discussion about the different contributions to the pressure appears in Sec. IV. The long-time limit of the complete problem is included in Sec. V and finally, in Sec. VI we include our final remarks and conclusions.

\section{The model and the field influence action}

In order to include effects of dissipation and noise (fluctuations) in the calculation of the Casimir force between two half-spaces of real material interacting with the EM field, we will develop a full CTP approach to the problem. Therefore, a microscopic physical model for the material of the bodies will be introduced. Once the model is defined, one expects  the material to  show effective macroscopic EM properties, that will  be contained in the Feynman-Vernon influence functional included in the CTP formalism.

\subsection{The Total System: EM Field plus Matter}

As in Ref. \cite{CTPGauge}, we will consider a composite system, consisting in two parts: the EM field $A^{\mu}$ (considered as a massless vector gauge field) and the real media, which is modeled as a continuous set of quantum Brownian particles representing the polarization density degree of freedom $\mathbf{P}$. These degrees of freedom (DOFs) are basically three-dimensional harmonic oscillators (with mass $M^{[j]}$ and frequency $\Omega^{[j]}$ in each direction) coupled to the field at each point of space in a realistic way. The composite system (EM field and DOFs) is also coupled to external baths of harmonic oscillators in each direction as in the well-known quantum Brownian motion theory (QBM). Therefore, the total action for the whole system is given by:

\begin{eqnarray}
S[A^{\mu},\mathbf{P}_{\mathbf{x}},\mathbf{q}_{n,\mathbf{x}}]&=&S_{0}[A^{\mu}]+S_{0}[\mathbf{P}_{\mathbf{x}}]+\sum_{n}S_{0}[\mathbf{q}_{n,\mathbf{x}}]\\
&&+~S_{\rm Curr}[A^{\mu},\mathbf{P}_{\mathbf{x}}]+\sum_{n}S_{\rm int}[\mathbf{P}_{\mathbf{x}},\mathbf{q}_{n,\mathbf{x}}],\nonumber
\end{eqnarray}
where the first three terms are the free actions for each part of the total system, while the interaction action between the field and the polarization DOFs is given by a current-type interaction which can be written in two equivalent ways by defining the conserved current four-vector as $J_{\mu}=(\nabla\cdot\mathbf{P},-\dot{\mathbf{P}})$ and the electric field $E^{j}=-\partial_{j}A^{0}-\partial_{0}A^{j}$:

\begin{equation}
S_{\rm Curr}[A^{\mu},\mathbf{P}_{\mathbf{x}}]=\lambda_{0}\left[\left(g~J_{\mu}\right)\ast A^{\mu}\right]=\lambda_{0}\left[\left(g~P^{j}\right)\ast E^{j}\right],
\end{equation}

\noindent where $A\ast B\equiv\int d^{4}x~A(\mathbf{x},t)~B(\mathbf{x},t)$ and $\lambda_{0}$ is the coupling constant between them. It is worth noting that,
when writing the interaction actions, we have introduced the matter distribution $g(\mathbf{x})$, which takes binary values (1 or 0) depending whether or not there is material at the spatial point $\mathbf{x}$.

On the other hand, the interaction between the polarization DOFs and each bath is simply linear coupling through the constants $\lambda_{n,\mathbf{x}}^{[j]}$ for each bath oscillator ($n$) in each point ($\mathbf{x}$) and direction ($j$):

\begin{equation}
S_{\rm int}[\mathbf{P}_{\mathbf{x}},\mathbf{q}_{n,\mathbf{x}}]=\int d\mathbf{x}~g(\mathbf{x})~\lambda_{n,\mathbf{x}}^{[j]}\left(P_{\mathbf{x}}^{j}\cdot q_{n,\mathbf{x}}^{j}\right),
\end{equation}
where $A\cdot B\equiv\int_{t_{i}}^{t_{f}} dt~A(t)~B(t)$, $\lambda_{n,\mathbf{x}}^{[j]}$ are coupling constants between the DOFs and the baths, and $[j]$ denotes that this superscript is not summed as in Einstein notation. It is clear that greek indices sum over $0,1,2,3$ while latin sum over spatial components only ($1,2,3$).

Finally, we will assume that the total system is initially uncorrelated, thus the initial density matrix is written as a direct product of each part, which we also assume to be initially in a thermal equilibrium at a proper characteristic temperatures ($\beta_{EM}, \beta_{\mathbf{P}_{\mathbf{x}}}, \beta_{B,\mathbf{x}}$- the material can also, in principle, be thermally inhomogeneous-),

\begin{equation}
\widehat{\rho}(t_{i})=\widehat{\rho}_{\rm EM}(t_{i})\otimes\widehat{\rho}_{\mathbf{P}_{\mathbf{x}}}(t_{i})\otimes\widehat{\rho}_{\{\mathbf{q}_{n,\mathbf{x}}\}}(t_{i}).
\end{equation}

\subsection{EM Field Influence Action}\label{SectEMFieldInfluenceAction}

As our goal is the study of  the dynamics of the EM field correlation functions, we start by coarse-graining the baths and the DOFs in order to obtain the exact EM field influence action. Since we have quadratic actions and we choose the initial states as Gaussian, the EM field influence action will have a quadratic form, with dissipation and noise kernels acting over the EM field.

It is well known from the QBM theory, coarse- graining the baths will result in QBM influence actions for each component of each DOF in each point of space, i. e., the result will be $S_{\rm IF}[P_{\mathbf{x}}^{j},P_{\mathbf{x}}^{'j}]$ including the QBM dissipation and noise kernels. Therefore, at this point, the EM field influence action is defined as:

\begin{eqnarray}
&&e^{iS_{\rm IF}[A^{\mu},A'^{\mu}]}=\int d\mathbf{P}_{\rm f}\int d\mathbf{P}_{\rm i}~d\mathbf{P}'_{\rm i}\int_{\mathbf{P}(t_{\rm i})=\mathbf{P}_{\rm i}}^{\mathbf{P}(t_{\rm f})=\mathbf{P}_{\rm f}}\mathcal{D}\mathbf{P}\times \nonumber\\
&&\int_{\mathbf{P}'(t_{\rm i})=\mathbf{P}'_{\rm i}}^{\mathbf{P}'(t_{\rm f})=\mathbf{P}_{\rm f}}\mathcal{D}\mathbf{P}'~e^{i\lambda_{0}\int d\mathbf{x}~g(\mathbf{x})\left(\nabla A^{0}\cdot\mathbf{P}+\mathbf{A}\cdot\dot{\mathbf{P}}-\nabla A'^{0}\cdot\mathbf{P}'-\mathbf{A}'\cdot\dot{\mathbf{P}}'\right)}\nonumber\\
&&\times~e^{i\left(S_{0}[\mathbf{P}]-S_{0}[\mathbf{P}']+S_{\rm IF}[\mathbf{P},\mathbf{P}']\right)}~\rho_{\mathbf{P}}\left(\mathbf{P}_{\rm i},\mathbf{P}'_{\rm i},t_{\rm i}\right).
\label{FieldInfluenceAction}
\end{eqnarray}

Following Ref.\cite{CTPGauge}, and considering an initial thermal state for the DOFs, the EM field influence action is given by:

\begin{eqnarray}
S_{\rm IF}[A^{\mu},A'^{\mu}]&=&\int d^{4}x\int d^{4}x'~\Delta A^{\mu}(x)\Big[-2~D_{\mu\nu}(x,x')\nonumber\\
&&\times~\Sigma A^{\nu}(x')+\frac{i}{2}~N_{\mu\nu}(x,x')~\Delta A^{\nu}(x')\Big],
\label{EMFieldInfluenceFunctional}
\end{eqnarray}
with $\Delta A^{\mu}=A^{'\mu}-A^{\mu}$, $\Sigma A^{\mu}=(A^{\mu}+A^{'\mu})/2$ and the dissipation and noise kernels are given,
respectively, by:

\begin{equation}
D_{\mu\nu}(x,x')=\Gamma_{\mu\nu}\,^{jk}~\mathbb{D}_{jk},
\label{EMDissipationKernel}
\end{equation}

\begin{equation}
N_{\mu\nu}(x,x')=\Gamma_{\mu\nu}\,^{jk}~\mathbb{N}_{jk},
\label{EMNoiseKernelBath}
\end{equation}
which ensures gauge invariance thanks to the fact that the differential operator $\Gamma_{\mu\nu}\,^{jk}\equiv\delta_{\mu}\,^{0}~\delta_{\nu}\,^{0}~\partial_{jk'}^{2}-\delta_{\mu}\,^{0}~\delta_{\nu}\,^{k}~\partial_{jt'}^{2}-\delta_{\mu}\,^{j}~\delta_{\nu}\,^{0}~\partial_{tk'}^{2}+\delta_{\mu}\,^{j}~\delta_{\nu}\,^{k}~\partial_{tt'}^{2}$ satisfies $\partial^{\mu}\Gamma_{\mu\nu}^{~~~jk}=\partial^{'\nu}\Gamma_{\mu\nu}^{~~~jk}\equiv 0$.

It is worth noting that the noise kernel presents two contributions, one associated to the DOFs and the other one associated to the baths: $\mathbb{N}_{jk}\equiv \mathbb{N}_{jk}^{P}+\mathbb{N}_{jk}^{B}$.

All in all, the dissipation and noise kernels are given by:

\begin{equation}
\mathbb{D}_{jk}(x,x')=\delta_{jk}~\delta(\mathbf{x}-\mathbf{x}')~g(\mathbf{x})~\frac{\lambda_{0,\mathbf{x}}^{2}}{2}~G_{\rm Ret,\mathbf{x}}^{[j]}(t-t'),
\end{equation}

\begin{eqnarray}
\mathbb{N}_{jk}^{P}(x,x')&=&\delta_{jk}~\delta(\mathbf{x}-\mathbf{x}')~\frac{g(\mathbf{x})\lambda_{0,\mathbf{x}}^{2}M_{\mathbf{x}}^{[j]}}{2\Omega^{[j]}_{\mathbf{x}}}\coth\left(\frac{\beta_{P^{j}_{\mathbf{x}}}~\Omega^{[j]}_{\mathbf{x}}}{2}\right)\nonumber\\
&&\times\left[\dot{G}^{[j]}_{\rm Ret,\mathbf{x}}(t-t_{i})~\dot{G}^{[j]}_{\rm Ret,\mathbf{x}}(t'-t_{i})\right.\nonumber\\
&&\left.+~\Omega^{[j]2}_{\mathbf{x}}~G^{[j]}_{\rm Ret,\mathbf{x}}(t-t_{i})~G^{[j]}_{\rm Ret,\mathbf{x}}(t'-t_{i})\right],
\label{MatrixNoiseKernelDOFs}
\end{eqnarray}

\begin{equation}
\mathbb{N}^{B}_{jk}(x,x')=\delta_{jk}~\delta(\mathbf{x}-\mathbf{x}')~g(\mathbf{x})~\lambda_{0,\mathbf{x}}^{2}~G_{\rm Ret,\mathbf{x}}^{[j]}\cdot N_{\mathbf{x}}^{[j]}\cdot\left[G_{\rm Ret,\mathbf{x}}^{[j]}\right]^{T},
\label{MatrixNoiseKernelBaths}
\end{equation}
where we are considering materials that can be  inhomogeneous and anisotropic (birefringent) \cite{CTPGauge}.

$N_{\mathbf{x}}^{[j]}$ is the QBM noise kernel generated by the thermal bath over the DOF in the direction $j$ and located at $\mathbf{x}$. The Green functions $G_{\rm Ret,\mathbf{x}}^{[j]}$ are the retarded Green functions for each DOF, which satisfy $G_{\rm Ret,\mathbf{x}}^{[j]}(0)=0$ and $\dot{G}_{\rm Ret,\mathbf{x}}^{[j]}(0)=1$. In other words, they are the QBM retarded Green functions having $\mathbf{x}$ and $[j]$ as parameters. These Green fuctions are defined once the type of baths are chosen at each point of space and direction. It is implicit that the chosen directions defining the anisotropic properties are the Fresnel's principal axis' basis (see Ref.\cite{CTPGauge}) because, as we shall see, they will allow us to define a diagonal permittivity tensor in this basis. In fact, we are considering the same basis for every point filled with material.  This is completely general since it is clear, for example, that disjoint bodies can have different Fresnel's basis. To include these features, it is necessary to include changes of basis between all the present basis. For simplicity, this will be omitted and, in fact, for isotropic materials this is unnecessary since the permittivity tensor is proportional to the identity matrix.

In the Lifshitz problem there are two different  bodies interacting, which are assumed to be homogeneous and isotropic. Therefore a few simplifications arise. On the one hand, homogeneity implies that all the spatial subscripts reduce to two labels associated to each of the two bodies. Considering two parallel plates separated by a distance $l$, the homogeneity of each body means $\mathbf{x}\rightarrow L,R$ depending on the spatial point lays in the left or right body respectively. On the other hand, isotropy of the material in each body implies that superscripts $[j]$ should be omitted because there is no dependence with the direction. However, we will keep generality in the material properties for now.

\section{Casimir  Pressure in terms of EM correlation functions}

In the present section we will calculate the EM field CTP-generating functional and derive an expression for the Hadamard propagator.  We shall show how the Casimir pressure can be easily written in terms of this Hadamard propagator.

\subsection{The Hadamard Propagator and Electrodynamics in the Temporal Gauge}

Once we have calculated the EM field influence action generated by the material, we proceed to calculate the EM field generating functional, which is defined by \cite{CTPGauge}:

\begin{eqnarray}
&&Z_{\rm CTP}[J_{\mu},J'_{\mu}]=\int dA^{\mu}_{\rm f}\int dA^{\mu}_{\rm i}~dA'^{\mu}_{\rm i}\int_{A^{\mu}(t_{\rm i})=A^{\mu}_{\rm i}}^{A^{\mu}(t_{\rm f})=A^{\mu}_{\rm f}}\mathcal{D}A^{\mu}\nonumber\\
&&\int_{A'^{\mu}(t_{\rm i})=A'^{\mu}_{\rm i}}^{A'^{\mu}(t_{\rm f})=A^{\mu}_{\rm f}}\mathcal{D}A'^{\mu}~e^{i\left(J_{\mu}\ast~A^{\mu}-J'_{\mu}\ast~A'^{\mu}\right)}~e^{i\left(S_{0}[A^{\mu}]-S_{0}[A'^{\mu}]\right)}\nonumber\\
&&\times~e^{iS_{\rm IF}[A^{\mu},A'^{\mu}]}~\rho_{\rm EM}(A^{\mu}_{\rm i},A'^{\mu}_{\rm i},t_{\rm i}).
\label{GFEMSinFaddeevPopov}
\end{eqnarray}

These integrals can be performed using the Faddeev-Popov procedure (adapted to the CTP formalism),  to extract the redundant sums over paths on the same gauge class. In \cite{CTPGauge}, the gauge condition was introduced in the CTP-action as a typical gauge fixing term. Then, the CTP integral is worked out by writing the paths as a sum of a homogeneous solution $A_{0}^{\mu}(x)$ (satisfying the initial conditions) plus a shift. The corresponding equations of motion for this solution are the ones obtained from the EM CTP action (including the EM field influence action), which in the case of the temporal gauge reads:

\begin{eqnarray}
\Big(\eta_{\mu\nu}\square-\partial_{\mu}\partial_{\nu}&-&\frac{1}{\alpha}~t_{\mu}t_{\nu}\Big)A^{\nu}(x)\label{EqMotionAmuAlpha}\\
&&+~2\int d^{4}x'~D_{\mu\nu}(x,x')~A^{\nu}(x')=0,\nonumber
\end{eqnarray}

\noindent where $\alpha$ is the gauge fixing parameter and $t^{\mu}$ is a time-like four-vector which can be taken as $(1,0,0,0)$ for the temporal gauge. These are four equations for the four components of the EM field.

Therefore, the functional integral can be calculated in the Landau gauge (where the gauge fixing parameter $\alpha$ goes to 0). All in all, the generating functional reads:

\begin{eqnarray}
Z_{\rm CTP}[\Sigma \mathbf{J},\Delta \mathbf{J}]&=&\Big\langle~e^{-i\Delta\mathbf{J}\ast\mathbf{A}_{0}}\Big\rangle_{\Sigma\mathbf{A}_{\rm i},\Sigma\mathbf{\Pi}_{\rm i}}~e^{-i\Delta\mathbf{J}\ast\mathcal{G}_{\rm Ret}\ast\Sigma\mathbf{J}}\nonumber\\
&&\times~e^{-\frac{1}{2}\Delta \mathbf{J}\ast\mathcal{G}_{\rm Ret}\ast \left(\partial_{tt'}^{2}\mathbb{N}\right)\ast\mathcal{G}_{\rm Ret}^{T}\ast\Delta \mathbf{J}}~,
\label{ZCTPGaugeTemporalFINAL}
\end{eqnarray}

\noindent where $\Big\langle...\Big\rangle_{\Sigma\mathbf{A}_{\rm i},\Sigma\mathbf{\Pi}_{\rm i}}=\int d\Sigma\mathbf{A}_{i}\int d\Sigma\mathbf{\Pi}_{i}~...W_{\rm EM}[\Sigma\mathbf{A}_{i},\Sigma\mathbf{\Pi}_{i},$ $t_{i}]$,  and $W_{\rm EM}\left[\Sigma\mathbf{A}_{i},\Sigma\mathbf{\Pi}_{i},t_{i}\right]$ is the initial EM field Wigner functional associated to the EM field initial density matrix $\widehat{\rho}_{\rm EM}(t_{i})$ in the temporal gauge (see Ref.\cite{CTPGauge} for its definition).

It is worth noting that $\mathbf{A}_{0}$ corresponds to the homogeneous solution of the EM equations of motion after imposing the temporal gauge, which is written as:

\begin{eqnarray}
A_{0}^{j}(x)&=&-\int d\mathbf{x}'~\dot{\mathcal{G}}_{\rm Ret}^{jk}(\mathbf{x},\mathbf{x}',t-t_{i})~\Sigma A_{\rm i}^{k}(\mathbf{x}')\nonumber\\
&&+\int d\mathbf{x}'~\mathcal{G}_{\rm Ret}^{jk}(\mathbf{x},\mathbf{x}',t-t_{i})~\Sigma\Pi_{\rm i}^{k}(\mathbf{x}').
\label{SolutionTemporalGauge}
\end{eqnarray}

$\mathcal{G}_{\rm Ret}$ corresponds to the retarded Green tensor resulting from Eq.(\ref{EqMotionAmuAlpha}) after imposing the temporal gauge condition in the Landau gauge, which results in a set of three equations of motion plus a residual gauge condition obtained from the equation for $A^{0}$ (which was erased by the gauge condition). The initial conditions for the retarded propagator are, as usual,

\begin{equation}
\mathcal{G}_{\rm Ret}^{jk}(\mathbf{x},\mathbf{x}',0)=0~~~,~~~~\dot{\mathcal{G}}_{\rm Ret}^{jk}(\mathbf{x},\mathbf{x}',0)=-~\delta^{jk}~\delta(\mathbf{x}-\mathbf{x}').
\label{GreenTensorInitialConditions}
\end{equation}

Once we have calculated the EM field generating functional, we can functionally derive  it in order to obtain the EM $n-$point functions and, in particular, the propagators.

The calculation of the Wightman function is straightforward, from which we can read the Hadamard propagator:

\begin{eqnarray}
\mathcal{G}_{\rm H}^{jk}(x_{1},x_{2})&\equiv&\Big\langle A_{0}^{j}(x_{1})A_{0}^{k}(x_{2})\Big\rangle_{\Sigma\mathbf{A}_{\rm i},\Sigma\mathbf{\Pi}_{\rm i}}\label{HadamardPropagator}\\
&&+\Big[\mathcal{G}_{\rm Ret}\ast \left(\partial_{tt'}^{2}\mathbb{N}\right)\ast\left(\mathcal{G}_{\rm Ret}\right)^{T}\Big]^{jk}(x_{1},x_{2}).\nonumber
\end{eqnarray}

This expression holds for every initial state of the field and depends on the chosen gauge. Note that the Hadamard propagator has two separated contributions. The first term is entirely associated to the field's effective dynamics and the initial state. The other contribution is associated to the material degrees of freedom represented by the noise kernel $\mathbb{N}$, which also splits in two contributions due to the composite nature of the material (DOFs plus bath in each point of space).

However, all the dynamics is up to the retarded Green tensor in the temporal gauge. As it was shown in Ref.\cite{CTPGauge}, the choice of the Landau gauge naturally implies that $A^{0}\equiv 0$ in order to avoid divergent terms. Considering the temporal component of Eq.(\ref{EqMotionAmuAlpha}), and imposing the gauge condition, one obtains a residual condition over the remaining components of the EM field, which can be written as:

\begin{equation}
\nabla\cdot\left[\int_{t_{i}}^{t}dt'~\partial_{t}\Big(\overleftrightarrow{\varepsilon}(t-t',\mathbf{x})\Big)\cdot\mathbf{A}(\mathbf{x},t')\right]=0,
\label{ResidualCondition}
\end{equation}

\noindent where the permittivity tensor for the inhomogeneous and anisotropic material is given by:

\begin{equation}
\varepsilon_{mr}(t-t',\mathbf{x})\equiv\delta_{mr}\left(\delta(t-t')+\lambda_{0,\mathbf{x}}^{2}~g(\mathbf{x})~G_{\rm Ret,\mathbf{x}}^{[m]}(t-t')\right).
\label{PermittivityTensor}
\end{equation}

Since the tensor is diagonal, the expression is given in the Fresnel's principal axes basis. The residual condition in Eq.(\ref{ResidualCondition}) is a generalization of the condition considered in Ref.\cite{EberleinRobaschik2006}. It was shown in Ref.\cite{CTPGauge} that in the case of isotropic and non-dissipative material, Eq.(\ref{ResidualCondition}) reduces to the generalized Coulomb condition of Ref.\cite{EberleinRobaschik2006} $$\nabla\cdot\left[\varepsilon(\mathbf{x})~\mathbf{A}(\mathbf{x},t)\right]=0,$$
where the permittivity tensor has been replaced by a single  function.

Also in Ref.\cite{CTPGauge} it is proved that, in the general case, the equations of motion for the spatial components of the EM field in the temporal gauge can be written as:

\begin{eqnarray}
\frac{\partial^{2}\mathbf{A}}{\partial t^{2}}&+&\nabla\times\left(\nabla\times\mathbf{A}\right)+\lambda_{0,\mathbf{x}}^{2}~g(\mathbf{x})~\mathbf{A}(\mathbf{x},t)\label{EqMotionAmTemporalGauge}\\
&&+~\lambda_{0,\mathbf{x}}^{2}~g(\mathbf{x})\int_{t_{i}}^{t}dt'~\ddot{\overleftrightarrow{\mathbf{G}}}_{\rm Ret,\mathbf{x}}(t-t')\cdot\mathbf{A}(\mathbf{x},t')=0,\nonumber
\end{eqnarray}
where $\left(\overleftrightarrow{\mathbf{G}}_{\rm Ret,\mathbf{x}}\right)_{mk}=\delta_{mk}~G_{\rm Ret,\mathbf{x}}^{[m]}$. Again, from the fact that we could write this diagonal tensor (associated to the retarded Green functions), it is clear that the chosen basis is the Fresnel's principal axes basis (in general, the tensor would be non-diagonal). It is also remarkable the appearance of the third term, which constitutes a finite renormalization position-dependent mass term for the EM field as the one found in the scalar case in Ref.\cite{CTPScalar}. As we shall see, this term will be irrelevant in the determination of the Green tensor.


Considering the equation of motion for the EM field, the retarded Green tensor $\overleftrightarrow{\mathcal{G}}_{\rm Ret}(\mathbf{x},\mathbf{x}',t)$ can be defined as:

\begin{eqnarray}
&&0=\frac{\partial^{2}\overleftrightarrow{\mathcal{G}}_{\rm Ret}}{\partial t^{2}}+\nabla\times\nabla\times\overleftrightarrow{\mathcal{G}}_{\rm Ret}+\lambda_{0,\mathbf{x}}^{2}g(\mathbf{x})\overleftrightarrow{\mathcal{G}}_{\rm Ret}(\mathbf{x},\mathbf{x}',t)\nonumber\\
&&+\lambda_{0,\mathbf{x}}^{2}g(\mathbf{x})\int_{0}^{t}dt'~\ddot{\overleftrightarrow{\mathbf{G}}}_{\rm Ret,\mathbf{x}}(t-t')\cdot\overleftrightarrow{\mathcal{G}}_{\rm Ret}(\mathbf{x},\mathbf{x}',t'),
\label{EqMotionEMRetGreenTensor}
\end{eqnarray}
with the initial conditions given in Eq.(\ref{GreenTensorInitialConditions}).

Laplace-transforming the equation, we easily obtain:

\begin{equation}
\nabla\times\nabla\times\overleftrightarrow{\mathcal{G}}_{\rm Ret}+s^{2}\overleftrightarrow{\varepsilon}(s,\mathbf{x})\cdot\overleftrightarrow{\mathcal{G}}_{\rm Ret}(\mathbf{x},\mathbf{x}',s)=-\mathbb{I}~\delta(\mathbf{x}-\mathbf{x}'),
\label{EqMotionEMRetGreenTensorLaplace}
\end{equation}
where the Laplace transform of the permittivity tensor of Eq.(\ref{PermittivityTensor}) is $$\overleftrightarrow{\varepsilon}(s,\mathbf{x})=\mathbb{I}+\lambda_{0,\mathbf{x}}^{2}~g(\mathbf{x})~\overleftrightarrow{\mathbf{G}}_{\rm Ret,\mathbf{x}}(s)\, .$$

It is clear that including dissipation it is not possible to define refractive indexes in the time domain. However, since the Laplace transform of the permittivity tensor turns out to be diagonal in this basis, the refractive indexes can be defined in the Laplace variable's domain. In each direction $j$, we can define the complex refractive indexes $$n_{\mathbf{x}}^{[j]2}=1+\lambda_{0,\mathbf{x}}^{2}~g(\mathbf{x})~G_{\rm Ret,\mathbf{x}}^{[j]}(s)$$ in such a way that the Fresnel's ellipsoid is a useful picture to describe the material's anisotropy.


From Eq.(\ref{EqMotionEMRetGreenTensorLaplace}) we can see that the Laplace transform of the EM retarded Green tensor turns out to be the Feynman's Green tensor associated to the differential operator $\nabla\times\nabla\times+~s^{2}~\overleftrightarrow{\varepsilon}(s,\mathbf{x})\cdot$, and therefore satisfies $\mathcal{G}_{\rm Ret}^{ij}(\mathbf{x},\mathbf{x}',s)=\mathcal{G}_{\rm Ret}^{ji}(\mathbf{x}',\mathbf{x},s)$. It is also worth to note that the reality of the EM retarded Green tensor in the time domain implies that$$\mathcal{G}_{\rm Ret}^{ij}(\mathbf{x},\mathbf{x}',s)=\mathcal{G}_{\rm Ret}^{*ij}(\mathbf{x},\mathbf{x}',s^{*}).$$

\subsection{Transient and Steady Pressure}\label{EMCFP-TSP}

We would like to write the pressure in terms of the Hadam- ard propagator. In Ref.\cite{CTPGauge}, an expression for the expectation values of the components of the  Maxwell tensor is given in terms of the Hadamard propagator,  using the point-splitting technique.  As we are interested in the calculation of the Casimir force in the Lifshitz problem,  a simpler expression than the one in Ref.\cite{CTPGauge} can be achieved.

Considering the symmetry of the configuration, the force between the bodies will be given only by the pressure in the perpendicular direction to the surfaces, i. e., along the direction parallel to the separation distance $l$, which we will call the $z$ axis. Therefore, in the Lifshitz problem, the pressure will be given directly by the $zz-$component of the Maxwell tensor, which can be written for a field point $x_{1}$ inside the gap as in Ref.\cite{Antezza}:

\begin{equation}
\widehat{T}^{zz}(x_{1})=-\frac{\Lambda^{ij}}{8\pi}\left[\widehat{E}^{i}(x_{1})~\widehat{E}^{j}(x_{1})+\widehat{B}^{i}(x_{1})~\widehat{B}^{j}(x_{1})\right],
\end{equation}

\noindent where the electric field is given by $E^{i}=-\partial_{0}A^{j}$, while the magnetic field is $B^{i}=\left(\nabla\times\mathbf{A}\right)^{i}$.  $\Lambda^{ij}$ is the diagonal matrix $\Lambda^{11}=\Lambda^{22}=1=-\Lambda^{33}$.

Then, using the point-splitting technique and typical relations between the different propagators (see \cite{CalHu}), the expectation value of the $zz-$component, can be written in terms of the Hadamard propagator:

\begin{eqnarray}
P_{\rm Cas}(x_{1})&\equiv&\left\langle\widehat{T}^{zz}(x_{1})\right\rangle=-\frac{\Lambda^{ij}}{8\pi}\lim_{x_{2}\rightarrow x_{1}}\Big[\delta^{is}~\delta^{jm}~\partial_{t_{1}}\partial_{t_{2}}\nonumber\\
&&+~\epsilon^{irs}\epsilon^{jlm}~\partial_{r_{1}}\partial_{l_{2}}\Big]\mathcal{G}_{\rm H}^{sm}(x_{1},x_{2}),
\label{PCas}
\end{eqnarray}

\noindent once the propagator is renormalized in the coincidence limit.

It is worth noting that, unlike the one found in Ref.\cite{Antezza} which is proposed to correspond to the steady situation, the last expression actually is the Casimir pressure which emerges at the initial time $t_{i}$. Therefore, it comprises all the transient dynamics of the pressure in the way to reach its final value at the steady situation. The pressure can depend on time but also on space during the transient stage, until finally achieves the steady situation, where its value results to be time and space-independent:

\begin{equation}
P_{\rm Cas}(x_{1})\rightarrow P_{\rm Cas}^{\infty}.
\end{equation}

As a final remark, given the splitting of the contributions to the Hadamard propagator in Eq.(\ref{HadamardPropagator}),  and also that $\mathbb{N}=\mathbb{N}^{P}+\mathbb{N}^{B}$ (depending each one on its temperature), it is clear that the total pressure can be written in terms of three contributions:

\begin{equation}
P_{\rm Cas}(x_{1})=P_{\rm IC}(x_{1})+P_{\rm DOFs}(x_{1},\beta_{P_{\mathbf{x}}})+P_{\rm B}(x_{1},\beta_{B,\mathbf{x}}).
\label{ContribucionesPresion}
\end{equation}

Then, each part of the total system will contribute to the Casimir pressure at a given space-time point. The main subject of study of the next sections will be to determine which contributions will survive in the steady situation.

As we are dealing with an initial conditions problem, every time variable is defined in the interval $[t_{i},+\infty)$. Therefore, we can Laplace transform in each time variable $t_{1},t_{2}$ inside the coincidence limit of Eq.(\ref{PCas}). Introducing a Mellin's formula for each variable, the second term in Eq.(\ref{PCas}) is easily written in terms of the retarded Green tensor's double Laplace transform (note that this term only involves spatial derivatives):

\begin{eqnarray}
&&\lim_{x_{2}\rightarrow x_{1}}\epsilon^{irs}\epsilon^{jlm}~\partial_{r_{1}}\partial_{l_{2}}\mathcal{G}_{\rm H}^{sm}(x_{1},x_{2})=\label{MagneticPressureLaplace}\\
&&=\int_{\alpha_{1}-i\infty}^{\alpha_{1}+i\infty}\frac{ds_{1}}{2\pi i}\int_{\alpha_{2}-i\infty}^{\alpha_{2}+i\infty}\frac{ds_{2}}{2\pi i}~e^{(s_{1}+s_{2})(t_{1}-t_{i})}\nonumber\\
&&\times\lim_{\mathbf{x}_{2}\rightarrow \mathbf{x}_{1}}\Big[\epsilon^{irs}\epsilon^{jlm}~\partial_{r_{1}}\partial_{l_{2}}\mathcal{G}_{\rm H}^{sm}(\mathbf{x}_{1},s_{1};\mathbf{x}_{2},s_{2})\Big],\nonumber
\end{eqnarray}

\noindent where in the r.h.s. the coincidence limit was taken for the time variables and where $\alpha_{1,2}$ are taken to define vertical lines in the $s_{1,2}-$complex planes in such a way that all the poles of the integrands taken as functions of $s_{1}$ and $s_{2}$ are at the left of these lines.

Writing each time variable of the first term of Eq.(\ref{PCas}) in terms of Laplace transforms, we find:

\begin{eqnarray}
&&\partial_{t_{1}}\partial_{t_{2}}\mathcal{G}_{\rm H}^{sm}(x_{1},x_{2})=\nonumber\\
&&=\int_{\alpha_{1}-i\infty}^{\alpha_{1}+i\infty}\frac{ds_{1}}{2\pi i}~e^{s_{1}(t_{1}-t_{i})}~\mathfrak{L}_{1}\Big[\partial_{t_{1}}\partial_{t_{2}}\mathcal{G}_{\rm H}^{sm}(x_{1},x_{2})\Big]\nonumber\\
&&=\int_{\alpha_{1}-i\infty}^{\alpha_{1}+i\infty}\frac{ds_{1}}{2\pi i}~e^{s_{1}(t_{1}-t_{i})}\Big[s_{1}\partial_{t_{2}}\mathcal{G}_{\rm H}^{sm}(\mathbf{x}_{1},s_{1};x_{2})\nonumber\\
&&~-~\partial_{t_{2}}\mathcal{G}_{\rm H}^{sm}(\mathbf{x}_{1},t_{i};x_{2})\Big]\nonumber\\
&&=\int_{\alpha_{1}-i\infty}^{\alpha_{1}+i\infty}\frac{ds_{1}}{2\pi i}\int_{\alpha_{2}-i\infty}^{\alpha_{2}+i\infty}\frac{ds_{2}}{2\pi i}~e^{s_{1}(t_{1}-t_{i})}~e^{s_{2}(t_{2}-t_{i})}\nonumber\\
&&\times\Big[s_{1}\mathfrak{L}_{2}\Big(\partial_{t_{2}}\mathcal{G}_{\rm H}^{sm}(\mathbf{x}_{1},s_{1};x_{2})\Big)-\mathfrak{L}_{2}\Big(\partial_{t_{2}}\mathcal{G}_{\rm H}^{sm}(\mathbf{x}_{1},t_{i};x_{2})\Big)\Big]\nonumber\\
&&=\int_{\alpha_{1}-i\infty}^{\alpha_{1}+i\infty}\frac{ds_{1}}{2\pi i}\int_{\alpha_{2}-i\infty}^{\alpha_{2}+i\infty}\frac{ds_{2}}{2\pi i}~e^{s_{1}(t_{1}-t_{i})}~e^{s_{2}(t_{2}-t_{i})}\nonumber\\
&&\times\Big[s_{1}s_{2}~\mathcal{G}_{\rm H}^{sm}(\mathbf{x}_{1},s_{1};\mathbf{x}_{2},s_{2})-s_{1}~\mathcal{G}_{\rm H}^{sm}(\mathbf{x}_{1},s_{1};\mathbf{x}_{2},t_{i})\nonumber\\
&&~-~s_{2}~\mathcal{G}_{\rm H}^{sm}(\mathbf{x}_{1},t_{i};\mathbf{x}_{2},s_{2})+\mathcal{G}_{\rm H}^{sm}(\mathbf{x}_{1},t_{i};\mathbf{x}_{2},t_{i})\Big].
\label{ElectricPressureLaplace}
\end{eqnarray}

The second term between brackets does not depend on $s_{2}$. Therefore, for that term, the integral over $s_{2}$ have the integrand $e^{s_{2}(t_{2}-t_{i})}$, which is analytic in all the $s_{2}-$complex plane, so the integral over any contour vanishes for $t_{2}>t_{i}$. The same happens for the third and last terms between brackets. All in all, we have proved that for the first term in Eq.(\ref{PCas}):

\begin{eqnarray}
&&\lim_{x_{2}\rightarrow x_{1}}\delta^{is}~\delta^{jm}~\partial_{t_{1}}\partial_{t_{2}}\mathcal{G}_{\rm H}^{sm}(x_{1},x_{2})=\nonumber\\
&&=\int_{\alpha_{1}-i\infty}^{\alpha_{1}+i\infty}\frac{ds_{1}}{2\pi i}\int_{\alpha_{2}-i\infty}^{\alpha_{2}+i\infty}\frac{ds_{2}}{2\pi i}~e^{(s_{1}+s_{2})(t_{1}-t_{i})}~s_{1}s_{2}~\delta^{is}\delta^{jm}\nonumber\\
&&\times\lim_{\mathbf{x}_{2}\rightarrow\mathbf{x}_{1}}\mathcal{G}_{\rm H}^{sm}(\mathbf{x}_{1},s_{1};\mathbf{x}_{2},s_{2}).
\end{eqnarray}

Finally, the Casimir pressure can be re-written as:

\begin{eqnarray}
&&P_{\rm Cas}(x_{1})=\frac{-1}{8\pi}\int_{\alpha_{1}-i\infty}^{\alpha_{1}+i\infty}\frac{ds_{1}}{2\pi i}\int_{\alpha_{2}-i\infty}^{\alpha_{2}+i\infty}\frac{ds_{2}}{2\pi i}~e^{(s_{1}+s_{2})(t_{1}-t_{i})}\nonumber\\
&&\lim_{\mathbf{x}_{2}\rightarrow\mathbf{x}_{1}}\Big[\Theta^{sm}(s_{1},s_{2})~\mathcal{G}_{\rm H}^{sm}(\mathbf{x}_{1},s_{1};\mathbf{x}_{2},s_{2})\Big],
\label{PCasDoubleLaplace}
\end{eqnarray}

\noindent where we have defined the operator $$\Theta^{sm}(s_{1},s_{2})\equiv\Lambda^{ij}\Big(s_{1}s_{2}~\delta^{is}~\delta^{jm}+\epsilon^{irs}\epsilon^{jlm}~\partial_{r_{1}}\partial_{l_{2}}\Big).$$

Last expression for the Casimir pressure seems formally the same to the one found in Ref.\cite{Antezza}, however there are subtle differences all related to the statement of both problems. Eq.(\ref{PCasDoubleLaplace}) is not a steady-situation expression (as the one in Ref.\cite{Antezza}), but it also comprises all the information about the transient evolution of the $zz-$component and describes the building up of the Casimir pressure in the long time regime. In fact, in Ref.\cite{Antezza}, the pressure is calculated from the electric field correlation, which results to be proportional to a Dirac $\delta$-function in the frequencies' difference, imposed by the steady situation formalism (based on the fluctuation-dissipation theorem) in stochastic electrodynamics (SED). Thus, the double integration is automatically reduced to one integration by the source correlation. In the present work, we calculate the pressure from the Hadamard propagator, which is related to the EM quantum field correlation and, as we shall see in next sections, during the transient evolution it is not necessarily proportional to a Dirac $\delta$-function. Finally, all these points are also reflected in the definition of the $\Theta$ operator. In Ref.\cite{Antezza}, it only depends on one frequency variable which appears as a denominator due to the fact that the pressure is calculated from the electric field correlation. The present operator is a function of two Laplace variables which appears as multiplicative factor, since we are calculating the full time evolution of the pressure from the EM field correlation.

In the next sections, we will calculate all the contributions to the pressure and study its time evolution, starting from the expression in Eq.(\ref{PCasDoubleLaplace}).

\section{Different Contributions to the Casimir Pressure}

The formal expression for the Casimir pressure obtained in Eq.(\ref{PCasDoubleLaplace}), is written  in terms of the Hadamard propagator's double Laplace transform. From Eq.(\ref{HadamardPropagator}) it is clear that it has two separated contributions, one associated to the EM field's initial conditions (the first term) and the other one associated to the material (the second term). In fact, the noise kernel also splits into two parts ($\mathbb{N}=\mathbb{N}^{P}+\mathbb{N}^{B}$), one associated to the DOFs ($\mathbb{N}^{P}$) and the other one associated to the baths ($\mathbb{N}^{B}$). Then, the Hadamard propagator's double Laplace transform reads:

\begin{eqnarray}
&&\mathcal{G}_{\rm H}^{jk}(\mathbf{x}_{1},s_{1};\mathbf{x}_{2},s_{2})\equiv\Big\langle A_{0}^{j}(\mathbf{x}_{1},s_{1})A_{0}^{k}(\mathbf{x}_{2},s_{2})\Big\rangle_{\Sigma\mathbf{A}_{\rm i},\Sigma\mathbf{\Pi}_{\rm i}}\label{HadamardPropagatorLaplace}\\
&&+~\mathfrak{L}_{1,2}\Big[\Big(\mathcal{G}_{\rm Ret}\ast \left(\partial_{tt'}^{2}\mathbb{N}\right)\ast\left(\mathcal{G}_{\rm Ret}\right)^{T}\Big)^{jk}\Big](\mathbf{x}_{1},s_{1};\mathbf{x}_{2},s_{2}).\nonumber
\end{eqnarray}

\noindent Each contribution will be analyzed separately.

\subsection{Initial Conditions' Contribution}

Provided that the homogeneous solution is given by Eq. (\ref{SolutionTemporalGauge}) and the initial conditions for the Green tensor are given in Eq.(\ref{GreenTensorInitialConditions}), the Laplace transform of the homogeneous solution results:

\begin{equation}
A_{0}^{j}(\mathbf{x},s)=\int d\mathbf{x}'~\mathcal{G}_{\rm Ret}^{jl}(\mathbf{x},\mathbf{x}',s)\left(\Sigma\Pi_{\rm i}^{l}(\mathbf{x}')-s~\Sigma A_{\rm i}^{l}(\mathbf{x}')\right),
\label{SolutionTemporalGaugeLaplace}
\end{equation}

\noindent where $\mathcal{G}_{\rm Ret}^{jl}(\mathbf{x},\mathbf{x}',s)$ is the Laplace transform of the EM retarded Green tensor.

Therefore, the first term of the double Laplace transform of the Hadamard propagator in Eq.(\ref{HadamardPropagatorLaplace}) results:

\begin{eqnarray}
\Big\langle A_{0}^{j}(\mathbf{x}_{1},s_{1})A_{0}^{k}(\mathbf{x}_{2},s_{2})\Big\rangle_{\Sigma\mathbf{A}_{\rm i},\Sigma\mathbf{\Pi}_{\rm i}}=\label{CorrelacionA0A0Laplace}~~~~~~~~~~~~~~~~~~~~~~~~~~~~\\
=\int d\mathbf{x}'\int d\mathbf{x}''~\mathcal{G}_{\rm Ret}^{jl}(\mathbf{x}_{1},\mathbf{x}',s_{1})~\mathcal{G}_{\rm Ret}^{km}(\mathbf{x}_{2},\mathbf{x}'',s_{2})~~~~~~~~~~~~\nonumber\\
\times\Big\langle\Big[\Sigma\Pi_{\rm i}^{l}-s_{1}\Sigma A_{\rm i}^{l}\Big](\mathbf{x}')\Big[\Sigma\Pi_{\rm i}^{m}-s_{2}\Sigma A_{\rm i}^{m}\Big](\mathbf{x}'')\Big\rangle_{\Sigma\mathbf{A}_{\rm i},\Sigma\mathbf{\Pi}_{\rm i}}.\nonumber
\end{eqnarray}

It is clear that calculating the brackets average over initial configurations will introduce the EM field's initial state through its Wigner functional. Therefore, this contribution clearly depends on the initial state for the EM field. However, instead of proceeding directly to the calculation of the average, it is more convenient to calculate the brackets as a quantum expectation value. As the field is free at the initial time $t_{i}$,  the quantum expectation values in the operator formalism and these averages (using the homogeneous solution) are closely related. In fact, as a particular case of the theory developed in previous sections, it is easy to show that a free theory (without interactions) verifies:

\begin{eqnarray}
&&\Big\langle\Big[\Sigma\Pi_{\rm i}^{l}-s_{1}\Sigma A_{\rm i}^{l}\Big](\mathbf{x}')\Big[\Sigma\Pi_{\rm i}^{m}-s_{2}\Sigma A_{\rm i}^{m}\Big](\mathbf{x}'')\Big\rangle_{\Sigma\mathbf{A}_{\rm i},\Sigma\mathbf{\Pi}_{\rm i}}\equiv\nonumber\\
&&\equiv\frac{1}{2}\Big\langle\Big\{\widehat{\Pi}_{\rm i}^{l}(\mathbf{x}')-s_{1}~\widehat{A}_{\rm i}^{l}(\mathbf{x}');\widehat{\Pi}_{\rm i}^{m}(\mathbf{x}'')-s_{2}~\widehat{A}_{\rm i}^{m}(\mathbf{x}'')\Big\}\Big\rangle,
\label{RelacionCorrelacionCasoLibre}
\end{eqnarray}

\noindent where the r.h.s. corresponds to the quantum expectation value of the anticommutator containing quantum free EM field operators in the temporal gauge. In the free EM field case, the temporal gauge also implies the Coulomb gauge condition ($\nabla\cdot\mathbf{A}=0$) over the remaining components. Therefore, the EM free field operators are:

\begin{eqnarray}
\widehat{A}_{i}^{j}(\mathbf{x})&=&\int\frac{d\mathbf{k}}{\sqrt{2\omega_{\mathbf{k}}(2\pi)^{3}}}\sum_{\lambda}\varepsilon^{j}(\mathbf{k},\lambda)\Big[\widehat{a}_{\mathbf{k},\lambda}~e^{-i\left(\omega_{\mathbf{k}}t_{i}-\mathbf{k}\cdot\mathbf{x}\right)}\nonumber\\
&&+~\widehat{a}_{\mathbf{k},\lambda}^{\dag}~e^{i\left(\omega_{\mathbf{k}}t_{i}-\mathbf{k}\cdot\mathbf{x}\right)}\Big],
\end{eqnarray}

\begin{eqnarray}
\widehat{\Pi}_{i}^{j}(\mathbf{x})&=&\int\frac{d\mathbf{k}}{\sqrt{2\omega_{\mathbf{k}}(2\pi)^{3}}}\sum_{\lambda}i\omega_{\mathbf{k}}\varepsilon^{j}(\mathbf{k},\lambda)\Big[\widehat{a}_{\mathbf{k},\lambda}e^{-i\left(\omega_{\mathbf{k}}t_{i}-\mathbf{k}\cdot\mathbf{x}\right)}\nonumber\\
&&-~\widehat{a}_{\mathbf{k},\lambda}^{\dag}~e^{i\left(\omega_{\mathbf{k}}t_{i}-\mathbf{k}\cdot\mathbf{x}\right)}\Big],
\end{eqnarray}

\noindent where $\lambda$ sums over the transverse electric (TE) and magnetic (TM) polarizations, $\widehat{a}_{\mathbf{k},\lambda},\widehat{a}_{\mathbf{k},\lambda}^{\dag}$ are the free EM field photon annihilation and creation operators, $\varepsilon^{j}(\mathbf{k},\lambda)$ is the $j-$component of the polarization vectors and $\omega_{\mathbf{k}}=|\mathbf{k}|$.

Considering the completeness relation for the polarization vectors:

\begin{equation}
\sum_{\lambda=\rm TE,TM}\varepsilon^{l}(\mathbf{k},\lambda)~\varepsilon^{m}(\mathbf{k},\lambda)=\delta^{lm}-\frac{k^{l}k^{m}}{\omega_{\mathbf{k}}^{2}},
\end{equation}

\noindent and an initial thermal state for the EM field we get

\begin{eqnarray}
\left\langle\widehat{a}_{\mathbf{k},\lambda}~\widehat{a}_{\mathbf{k}',\lambda'}^{\dag}\right\rangle=\delta_{\lambda\lambda'}~\delta(\mathbf{k}-\mathbf{k}')\left(1+N(\omega_{\mathbf{k}})\right),
\end{eqnarray}

\begin{eqnarray}
\left\langle\widehat{a}_{\mathbf{k},\lambda}^{\dag}~\widehat{a}_{\mathbf{k}',\lambda'}\right\rangle=\delta_{\lambda\lambda'}~\delta(\mathbf{k}-\mathbf{k}')~N(\omega_{\mathbf{k}}),
\end{eqnarray}

\begin{equation}
\Big\langle\widehat{a}_{\mathbf{k},\lambda}~\widehat{a}_{\mathbf{k}',\lambda'}\Big\rangle=\left\langle\widehat{a}_{\mathbf{k},\lambda}^{\dag}~\widehat{a}_{\mathbf{k}',\lambda'}^{\dag}\right\rangle=0,
\end{equation}

\noindent where $N(\omega_{\mathbf{k}})=\frac{1}{\left(e^{\beta_{\rm EM}\omega_{\mathbf{k}}}-1\right)}$ is the photon occupation number for the initial thermal state of temperature $\beta_{\rm EM}$. Then, after a change of variables, Eq.(\ref{RelacionCorrelacionCasoLibre}) can be written as:

\begin{eqnarray}
&&\Big\langle\Big[\Sigma\Pi_{\rm i}^{l}-s_{1}\Sigma A_{\rm i}^{l}\Big](\mathbf{x}')\Big[\Sigma\Pi_{\rm i}^{m}-s_{2}\Sigma A_{\rm i}^{m}\Big](\mathbf{x}'')\Big\rangle_{\Sigma\mathbf{A}_{\rm i},\Sigma\mathbf{\Pi}_{\rm i}}=\nonumber\\
&&=\int\frac{d\mathbf{k}}{2\omega_{\mathbf{k}}(2\pi)^{3}}\left(\delta^{lm}-\frac{k^{l}k^{m}}{\omega_{\mathbf{k}}^{2}}\right)\coth\left(\frac{\beta_{\rm EM}}{2}~\omega_{\mathbf{k}}\right)\times\nonumber\\
&&\times~e^{i\mathbf{k}\cdot(\mathbf{x}'-\mathbf{x}'')}\left(s_{1}s_{2}+\omega_{\mathbf{k}}^{2}\right).
\end{eqnarray}

Therefore, the average over the initial conditions in Eq.(\ref{CorrelacionA0A0Laplace}) results:

\begin{eqnarray}
&&\Big\langle A_{0}^{j}(\mathbf{x}_{1},s_{1})A_{0}^{k}(\mathbf{x}_{2},s_{2})\Big\rangle_{\Sigma\mathbf{A}_{\rm i},\Sigma\mathbf{\Pi}_{\rm i}}=\label{CorrelacionA0A0LaplaceEstadoTermico}\\
&&=\int\frac{d\mathbf{k}}{2\omega_{\mathbf{k}}(2\pi)^{3}}\left[\delta^{lm}-\frac{k^{l}k^{m}}{\omega_{\mathbf{k}}^{2}}\right]\coth\left[\frac{\beta_{\rm EM}\omega_{\mathbf{k}}}{2}\right]\left[s_{1}s_{2}+\omega_{\mathbf{k}}^{2}\right]\nonumber\\
&&~~\times\left(\int d\mathbf{x}'~\mathcal{G}_{\rm Ret}^{jl}(\mathbf{x}_{1},\mathbf{x}',s_{1})~e^{i\mathbf{k}\cdot\mathbf{x}'}\right)\nonumber\\
&&~~\times\left(\int d\mathbf{x}''~\mathcal{G}_{\rm Ret}^{km}(\mathbf{x}_{2},\mathbf{x}'',s_{2})~e^{-i\mathbf{k}\cdot\mathbf{x}''}\right).\nonumber
\end{eqnarray}

Having this result, we can obtain an expression for the contribution of the initial conditions to the Casimir pressure. Replacing $\mathcal{G}_{\rm H}^{sm}$ in Eq.(\ref{PCasDoubleLaplace}) by the initial conditions' contribution $\Big\langle A_{0}^{j}(\mathbf{x}_{1},s_{1})A_{0}^{k}(\mathbf{x}_{2},s_{2})\Big\rangle_{\Sigma\mathbf{A}_{\rm i},\Sigma\mathbf{\Pi}_{\rm i}}$ it is straightforward that:

\begin{eqnarray}
&&P_{\rm IC}(x_{1},\beta_{EM})=-\frac{1}{8\pi}\int\frac{d\mathbf{k}}{2\omega_{\mathbf{k}}(2\pi)^{3}}\left[\delta^{lm}-\frac{k^{l}k^{m}}{\omega_{\mathbf{k}}^{2}}\right]\nonumber\\
&&\times\coth\left[\frac{\beta_{\rm EM}\omega_{\mathbf{k}}}{2}\right]\int_{\alpha_{1}-i\infty}^{\alpha_{1}+i\infty}\frac{ds_{1}}{2\pi i}\int_{\alpha_{2}-i\infty}^{\alpha_{2}+i\infty}\frac{ds_{2}}{2\pi i}\nonumber\\
&&\times~e^{(s_{1}+s_{2})(t_{1}-t_{i})}\left(s_{1}s_{2}+\omega_{\mathbf{k}}^{2}\right)\lim_{\mathbf{x}_{2}\rightarrow\mathbf{x}_{1}}\Bigg[\Theta^{jk}(s_{1},s_{2})\nonumber\\
&&\times\left(\int d\mathbf{x}'~\mathcal{G}_{\rm Ret}^{jl}(\mathbf{x}_{1},\mathbf{x}',s_{1})~e^{i\mathbf{k}\cdot\mathbf{x}'}\right)\nonumber\\
&&\times\left(\int d\mathbf{x}''~\mathcal{G}_{\rm Ret}^{km}(\mathbf{x}_{2},\mathbf{x}'',s_{2})~e^{-i\mathbf{k}\cdot\mathbf{x}''}\right)\Bigg].
\label{PICDoubleLaplace}
\end{eqnarray}

Remarkably,  this expression is quite general. In fact, it comprises the full time evolution of the contribution of the initial conditions to the Casimir pressure at any time for any point of space in a vacuum region, since the boundaries appear at the initial time $t_{i}$. This is the reason why it is physically expected that the contribution depends not only in time but also in space until reaching the steady situation. The information about the specific time evolution of the problem is encoded in the analytical properties of the integrand as a function of $s_{1}$ and $s_{2}$, i.e. in the poles and the branch cuts present,  as we shall see below.

Moreover, it is important to note that this expression is valid for any geometry of the boundaries,  including at least one vacuum region  (where we calculate the pressure). The information about the boundaries is contained in the Laplace transforms of the retarded Green tensors which have to be calculated in a specific situation in order to obtain a complete result. In next sections, we will show how this works for the Lifshitz problem,  deriving the long-time limit of this contribution.

\subsection{Material's Contribution}

Let us  consider the second term in the r.h.s. of Eq.(\ref{HadamardPropagatorLaplace}), which is associated to the material. Due to the fact that $\mathbb{N}=\mathbb{N}^{P}+\mathbb{N}^{B}$,  the contribution splits into two contributions, one associated to the DOFs and the other one associated to the baths. However, the first step in Laplace-transforming the contribution is the same. As the contribution reads $\mathfrak{L}_{1,2}\Big[\Big(\mathcal{G}_{\rm Ret}\ast \left(\partial_{tt'}^{2}\mathbb{N}\right)\ast\left(\mathcal{G}_{\rm Ret}\right)^{T}\Big)^{jk}\Big](\mathbf{x}_{1},s_{1};\mathbf{x}_{2},$ $s_{2})$ and the retarded Green tensors depend on the time differences, i. e., $\mathcal{G}_{\rm Ret}^{ij}(x,x')=\mathcal{G}_{\rm Ret}^{ij}(\mathbf{x},\mathbf{x}',t-t')$, then the products $\ast$ involve convolutions in the time variables between the noise kernel $\partial_{tt'}^{2}\mathbb{N}$ and one of the retarded Green tensors. Laplace-transforming is straightforward:

\begin{eqnarray}
&&\mathfrak{L}_{1,2}\Big[\Big(\mathcal{G}_{\rm Ret}\ast \left(\partial_{tt'}^{2}\mathbb{N}\right)\ast\left(\mathcal{G}_{\rm Ret}\right)^{T}\Big)^{jk}\Big](\mathbf{x}_{1},s_{1};\mathbf{x}_{2},s_{2})=\nonumber\\
&&=\int d\mathbf{x}\int d\mathbf{x}'~\mathcal{G}_{\rm Ret}^{jl}(\mathbf{x}_{1},\mathbf{x},s_{1})~\mathfrak{L}_{1,2}\left[\partial_{tt'}^{2}\mathbb{N}_{lm}\right](\mathbf{x},s_{1};\mathbf{x}',s_{2})\nonumber\\
&&~~\times~\mathcal{G}_{\rm Ret}^{km}(\mathbf{x}_{2},\mathbf{x}',s_{2}).
\end{eqnarray}

A few simplifications arise by considering that the noise kernels satisfy
$$\partial_{tt'}^{2}\mathbb{N}_{lm}(x,x')=\delta_{lm}~\delta(\mathbf{x}-\mathbf{x}')~g(\mathbf{x})~\partial_{tt'}^{2}\mathbb{N}_{\mathbf{x}}^{[l]}(t,t')$$ (see Eqs.(\ref{MatrixNoiseKernelDOFs}) and (\ref{MatrixNoiseKernelBaths}) ). Therefore,  taking into account the $\delta$-functions:

\begin{eqnarray}
&&\mathfrak{L}_{1,2}\Big[\Big(\mathcal{G}_{\rm Ret}\ast \left(\partial_{tt'}^{2}\mathbb{N}\right)\ast\left(\mathcal{G}_{\rm Ret}\right)^{T}\Big)^{jk}\Big](\mathbf{x}_{1},s_{1};\mathbf{x}_{2},s_{2})=\nonumber\\
&&=\int d\mathbf{x}~g(\mathbf{x})~\mathcal{G}_{\rm Ret}^{jl}(\mathbf{x}_{1},\mathbf{x},s_{1})~\mathfrak{L}_{1,2}\left[\partial_{tt'}^{2}\mathbb{N}_{\mathbf{x}}^{[l]}\right](s_{1},s_{2})\nonumber\\
&&\times~\mathcal{G}_{\rm Ret}^{kl}(\mathbf{x}_{2},\mathbf{x},s_{2}),
\end{eqnarray}

\noindent where the presence of $g$ denotes the fact that this contribution comes from the material regions exclusively.

Considering the  definitions in Eqs.(\ref{MatrixNoiseKernelDOFs}) and (\ref{MatrixNoiseKernelBaths}), Lapla- ce-transforming in both time variables the EM noise kernel (taking into account that $\mathbb{N}_{\mathbf{x},\rm B}^{[l]}(t_{i},t_{i})=0=\mathbb{N}_{\mathbf{x},\rm B}^{[l]}(s_{1},t_{i})=\mathbb{N}_{\mathbf{x},\rm B}^{[l]}(t_{i},s_{2})$ due to the causality of $G_{\rm Ret,\mathbf{x}}^{[l]}$), we obtain:

\begin{eqnarray}
\mathfrak{L}_{1,2}\left[\partial_{tt'}^{2}\mathbb{N}_{\mathbf{x}}^{[l]}\right](s_{1},s_{2})&=&s_{1}s_{2}~\mathbb{N}_{\mathbf{x},\rm P}^{[l]}(s_{1},s_{2})-s_{1}~\mathbb{N}_{\mathbf{x},\rm P}^{[l]}(s_{1},t_{i})\nonumber\\
&-&s_{2}~\mathbb{N}_{\mathbf{x},\rm P}^{[l]}(t_{i},s_{2})+\mathbb{N}_{\mathbf{x},\rm P}^{[l]}(t_{i},t_{i})\nonumber\\
&+&s_{1}s_{2}~\mathbb{N}_{\mathbf{x},\rm B}^{[l]}(s_{1},s_{2}),
\label{LaplaceNoiseContribution}
\end{eqnarray}

\noindent where each Laplace transform is given by:

\begin{eqnarray}
&&\mathbb{N}_{\mathbf{x},\rm P}^{[l]}(s_{1},s_{2})=\frac{\lambda_{0,\mathbf{x}}^{2}M_{\mathbf{x}}^{[j]}}{2\Omega^{[j]}_{\mathbf{x}}}\coth\left(\frac{\beta_{P^{j}_{\mathbf{x}}}~\Omega^{[j]}_{\mathbf{x}}}{2}\right)\left(s_{1}s_{2}+\Omega_{\mathbf{x}}^{[l]2}\right)\nonumber\\
&&\times~G_{\rm Ret,\mathbf{x}}^{[l]}(s_{1})~G_{\rm Ret,\mathbf{x}}^{[l]}(s_{2}),
\end{eqnarray}

\begin{eqnarray}
\mathbb{N}_{\mathbf{x},\rm P}^{[l]}(s_{n},t_{i})&=&\mathbb{N}_{\mathbf{x},\rm P}^{[l]}(t_{i},s_{n})\\
&=&\frac{\lambda_{0,\mathbf{x}}^{2}M_{\mathbf{x}}^{[l]}}{2\Omega^{[l]}_{\mathbf{x}}}\coth\left(\frac{\beta_{P^{l}_{\mathbf{x}}}~\Omega^{[l]}_{\mathbf{x}}}{2}\right)~s_{n}~G_{\rm Ret,\mathbf{x}}^{[l]}(s_{n}),\nonumber
\end{eqnarray}

\begin{equation}
\mathbb{N}_{\mathbf{x},\rm P}^{[l]}(t_{i},t_{i})=\frac{\lambda_{0,\mathbf{x}}^{2}M_{\mathbf{x}}^{[l]}}{2\Omega^{[l]}_{\mathbf{x}}}\coth\left(\frac{\beta_{P^{l}_{\mathbf{x}}}~\Omega^{[l]}_{\mathbf{x}}}{2}\right),
\end{equation}

\begin{equation}
\mathbb{N}_{\mathbf{x},\rm B}^{[l]}(s_{1},s_{2})=\lambda_{0,\mathbf{x}}^{2}~G_{\rm Ret,\mathbf{x}}^{[l]}(s_{1})~N_{\mathbf{x}}^{[l]}(s_{1},s_{2})~G_{\rm Ret,\mathbf{x}}^{[l]}(s_{2}).
\label{NBS1S2SINQBM}
\end{equation}

The last term in the r.h.s. of Eq.(\ref{LaplaceNoiseContribution}) is the only one associated to the baths. Moreover, it is important to note that the second and third terms only depend on one of the Laplace variables, while the fourth term does not depend on them.

Is clear that in order to continue the explicit calculation, one has to define the type of baths which are considered in each direction and point of space for the specific problem. However, we can take a further step in the calculation without loosing generality. From the QBM theory, is clear that for any type of bath, the QBM noise depends on the time differences, i. e., in the time domain we have that $N_{\mathbf{x}}^{[l]}(t,t')=N_{\mathbf{x}}^{[l]}(t-t')$. Therefore, although each time variable is defined in the interval $[t_{i},+\infty)$, its difference results defined in $(-\infty,+\infty)$. Assuming that the kernel verifies the convergence requirements (this can be easily implemented by introducing a cut-off function in the QBM noise kernel), it can be written in terms of its Fourier transform:

\begin{equation}
N_{\mathbf{x}}^{[l]}(t-t')=\int_{-\infty}^{+\infty}\frac{d\omega}{2\pi}~e^{-i\omega(t-t')}~\overline{N}_{\mathbf{x}}^{[l]}(\omega),
\end{equation}

\noindent where  $\overline{N}_{\mathbf{x}}^{[l]}(\omega)$ contains the dependencies  on the temperatures of the baths $\beta_{\mathbf{x},\rm B}$.

Using this general form for the QBM noise kernel, its Laplace transform is given by Eq.(\ref{NBS1S2SINQBM}):

\begin{eqnarray}
\mathbb{N}_{\mathbf{x},\rm B}^{[l]}(s_{1},s_{2})&=&\lambda_{0,\mathbf{x}}^{2}~G_{\rm Ret,\mathbf{x}}^{[l]}(s_{1})~G_{\rm Ret,\mathbf{x}}^{[l]}(s_{2})\int_{-\infty}^{+\infty}\frac{d\omega}{2\pi}\nonumber\\
&&\times\frac{\overline{N}_{\mathbf{x}}^{[l]}(\omega)}{(s_{1}+i\omega)(s_{2}-i\omega)}.
\label{NBS1S2CONQBM}
\end{eqnarray}

Once we have calculated each term of Eq.(\ref{LaplaceNoiseContribution}), the Eq.(\ref{PCasDoubleLaplace}) gives expressions for each contribution to the Casi- mir pressure in Eq.(\ref{ContribucionesPresion}). Therefore, we can write:

\begin{eqnarray}
&&P_{\rm DOFs}(x_{1},\beta_{P_{\mathbf{x}}^{l}})=-\frac{1}{8\pi}\int_{\alpha_{1}-i\infty}^{\alpha_{1}+i\infty}\frac{ds_{1}}{2\pi i}\int_{\alpha_{2}-i\infty}^{\alpha_{2}+i\infty}\frac{ds_{2}}{2\pi i}\nonumber\\
&&\times~e^{(s_{1}+s_{2})(t_{1}-t_{i})}\lim_{\mathbf{x}_{2}\rightarrow\mathbf{x}_{1}}\Bigg[\Theta^{jk}(s_{1},s_{2})\int d\mathbf{x}~g(\mathbf{x})\frac{\lambda_{0,\mathbf{x}}^{2}M_{\mathbf{x}}^{[l]}}{2\Omega^{[l]}_{\mathbf{x}}}\nonumber\\
&&\times\coth\left[\frac{\beta_{P^{l}_{\mathbf{x}}}\Omega^{[l]}_{\mathbf{x}}}{2}\right]\Big(\left[s_{1}^{2}G_{\rm Ret,\mathbf{x}}^{[l]}(s_{1})-1\right]\left[s_{2}^{2}G_{\rm Ret,\mathbf{x}}^{[l]}(s_{2})-1\right]\nonumber\\
&&+~\Omega^{[l]2}_{\mathbf{x}}~s_{1}s_{2}~G_{\rm Ret,\mathbf{x}}^{[l]}(s_{1})~G_{\rm Ret,\mathbf{x}}^{[l]}(s_{2})\Big)\nonumber\\
&&\times~\mathcal{G}_{\rm Ret}^{jl}(\mathbf{x}_{1},\mathbf{x},s_{1})~\mathcal{G}_{\rm Ret}^{kl}(\mathbf{x}_{2},\mathbf{x},s_{2})\Bigg],
\label{PDOFDoubleLaplace}
\end{eqnarray}

\begin{eqnarray}
&&P_{\rm B}(x_{1},\beta_{\mathbf{x},\rm B})=-\frac{1}{8\pi}\int_{-\infty}^{+\infty}\frac{d\omega}{2\pi}\int_{\alpha_{1}-i\infty}^{\alpha_{1}+i\infty}\frac{ds_{1}}{2\pi i}\int_{\alpha_{2}-i\infty}^{\alpha_{2}+i\infty}\frac{ds_{2}}{2\pi i}\nonumber\\
&&\times\frac{s_{1}s_{2}~e^{(s_{1}+s_{2})(t_{1}-t_{i})}}{(s_{1}+i\omega)(s_{2}-i\omega)}\lim_{\mathbf{x}_{2}\rightarrow\mathbf{x}_{1}}\Big[\Theta^{jk}(s_{1},s_{2})\int d\mathbf{x}~g(\mathbf{x})~\lambda_{0,\mathbf{x}}^{2}\nonumber\\
&&\times~\mathcal{G}_{\rm Ret}^{jl}(\mathbf{x}_{1},\mathbf{x},s_{1})~G_{\rm Ret,\mathbf{x}}^{[l]}(s_{1})~G_{\rm Ret,\mathbf{x}}^{[l]}(s_{2})~\overline{N}_{\mathbf{x}}^{[l]}(\omega)\nonumber\\
&&\times~\mathcal{G}_{\rm Ret}^{kl}(\mathbf{x}_{2},\mathbf{x},s_{2})\Big].
\label{PBathDoubleLaplace}
\end{eqnarray}

\noindent Note that this expression includes an integral over the Fourier frequencies.

Last results are valid in the most general case, which is an inhomogeneous and anisotropic material having local initial temperatures at each point. However, these expressions are simplified when considering the interaction between $n$ different but homogeneous and isotropic material bodies of volume $V_{n}$,  at homogeneous temperatures. In this case, all the spatial subscripts $\mathbf{x}$ (denoting inhomogeneity) and superscripts $[l]$ (denoting anisotropy) have to be replaced by a single index $n$ denoting the material body. Therefore the integral over the space splits as a sum of integrals over each volume $V_{n}$, and both expressions can be written as:

\begin{eqnarray}
P_{\rm DOFs}(x_{1},\beta_{P_{n}})&=&-\frac{1}{8\pi}\sum_{n}\frac{\lambda_{0,n}^{2}M_{n}}{2\Omega_{n}}\coth\left(\frac{\beta_{P_{n}}\Omega_{n}}{2}\right)\nonumber\\
&\times&\int_{\alpha_{1}-i\infty}^{\alpha_{1}+i\infty}\frac{ds_{1}}{2\pi i}\int_{\alpha_{2}-i\infty}^{\alpha_{2}+i\infty}\frac{ds_{2}}{2\pi i}e^{(s_{1}+s_{2})(t_{1}-t_{i})}\nonumber\\
&\times&\Big[\left(s_{1}^{2}G_{\rm Ret,n}(s_{1})-1\right)\left(s_{2}^{2}G_{\rm Ret,n}(s_{2})-1\right)\nonumber\\
&+&\Omega^{2}_{n}~s_{1}s_{2}~G_{\rm Ret,n}(s_{1})~G_{\rm Ret,n}(s_{2})\Big]\nonumber\\
&\times&\lim_{\mathbf{x}_{2}\rightarrow\mathbf{x}_{1}}\Bigg[\Theta^{jk}(s_{1},s_{2})\int_{V_{n}} d\mathbf{x}\mathcal{G}_{\rm Ret}^{jl}(\mathbf{x}_{1},\mathbf{x},s_{1})\nonumber\\
&\times&\mathcal{G}_{\rm Ret}^{kl}(\mathbf{x}_{2},\mathbf{x},s_{2})\Bigg],
\label{PDOFDoubleLaplaceUniformBodies}
\end{eqnarray}

\begin{eqnarray}
&&P_{\rm B}(x_{1},\beta_{n,\rm B})=\frac{-1}{8\pi}\sum_{n}\lambda_{0,n}^{2}\int_{-\infty}^{+\infty}\frac{d\omega}{2\pi}\overline{N}_{n}(\omega)\int_{\alpha_{1}-i\infty}^{\alpha_{1}+i\infty}\frac{ds_{1}}{2\pi i}\nonumber\\
&&\times\int_{\alpha_{2}-i\infty}^{\alpha_{2}+i\infty}\frac{ds_{2}}{2\pi i}\frac{s_{1}s_{2}~e^{(s_{1}+s_{2})(t_{1}-t_{i})}}{(s_{1}+i\omega)(s_{2}-i\omega)}~G_{\rm Ret,n}(s_{1})~G_{\rm Ret,n}(s_{2})\nonumber\\
&&\times\lim_{\mathbf{x}_{2}\rightarrow\mathbf{x}_{1}}\Big[\Theta^{jk}(s_{1},s_{2})\int_{V_{n}}d\mathbf{x}~\mathcal{G}_{\rm Ret}^{jl}(\mathbf{x}_{1},\mathbf{x},s_{1})\nonumber\\
&&\times~\mathcal{G}_{\rm Ret}^{kl}(\mathbf{x}_{2},\mathbf{x},s_{2})\Big].
\label{PBathDoubleLaplaceUniformBodies}
\end{eqnarray}

In this case, the quantity $$\lim_{\mathbf{x}_{2}\rightarrow\mathbf{x}_{1}}\Big[\Theta^{jk}(s_{1},s_{2})\int_{V_{n}}d\mathbf{x} \mathcal{G}_{\rm Ret}^{jl}(\mathbf{x}_{1},\mathbf{x},s_{1})~\mathcal{G}_{\rm Ret}^{kl}(\mathbf{x}_{2},\mathbf{x},s_{2})\Big]$$ is common to both contributions. However, the Mellin's integrals are the ones that define the time evolution and steady state of these contributions, and are very different due to the analytical properties of each integrand.

Moreover, is important to have in mind that in these expressions, it is implicit that at every point of the material bodies, the basis of Fresnel principal axis is the same. Each point defines a Fresnel's ellipsoid having its axis along the directions in which each component of the DOFs fluctuates. This is, in principle, a limitation in the present model, although it could be easily fixed by considering different basis in each body (or even at each point of space). As we mentioned before, in order to simplify the calculations but without losing the anisotropy properties, we will keep an unique basis for all the material points. This limitation disappears when considering isotropic materials, as in the Lifshitz problem analyzed below.

However, it is important to recognize that the ellipsoid's picture is valid for real refractive indexes. In our case, as we shall see, due to the inclusion of dissipation in the problem, the three refractive indexes  are complex, albeit the Fresnel's ellipsoid results useful as a conceptual and practical picture.

As final remarks, on the one hand, it is worth noting that considering this case for the initial conditions' contribution has no formal simplification over Eq.(\ref{PICDoubleLaplace}) since the initial field modes and the spatial integrals are over the whole space ($g(\mathbf{x})$ does not appear in those integrals). On the other hand, it is important to take into account that all the expressions containing a coincidence limit have to be regularized before calculating the limit. For the case of the Lifshitz problem, this will be done after integrating over the space.

\section{The Long-Time Limit of the Full Problem}

With the general expressions for each contribution of the Casimir pressure generated by different bodies of homogeneous and isotropic material at different temperatures in a vacuum region, we can study the  Lifshitz problem consisting in two half-spaces by letting $n$ be $L$ or $R$ for the left and right plate respectively,  as we have mentioned at the end of Sect. \ref{SectEMFieldInfluenceAction}.

\subsection{Analytical Properties of the Green Functions for the Lifshitz Problem}

From Eqs.(\ref{PICDoubleLaplace}), (\ref{PDOFDoubleLaplaceUniformBodies}) and (\ref{PBathDoubleLaplaceUniformBodies}) the time evolution and steady state of each contribution will be governed by the analytical properties of the integrands as functions of the Laplace variables $s_{1},s_{2}$. Therefore, for a given problem is critical to know the analytical properties of both Laplace transforms of the QBM's retarded Green function $G_{\rm Ret,n}(s)$ and EM retarded Green tensor $\mathcal{G}_{\rm Ret}^{ij}(\mathbf{x},\mathbf{x}',s)$.

As the DOFs are considered as quantum Brownian particles, irrespective of what type of environment is considered  the Laplace transform of the retarded Green function can be easily obtained from the equation of motion of the QBM's theory. In the present case, the result is the same as the one found in Ref.\cite{CTPScalar}. Then, we can easily write:

\begin{equation}
G_{\rm Ret,n}(s)=\frac{1}{\Big(s^{2}+\Omega_{n}^{2}-2~D_{n}(s)\Big)},
\label{QBMRetardedGreenFunction}
\end{equation}

\noindent where $D_{n}(s)$ is the Laplace transform of the QBM's retarded Green function for the environment (baths) of the plate $n$.

Thus, given a spectral density for the baths, the location of the poles will define the time evolution and the asymptotic behavior of the retarded Green function. However, causality implies, by Cauchy's theorem, that the poles of $G_{\rm Ret,n}$ should be located in the left-half of the complex $s-$plane. In fact, assuming that $\Omega_{n}\neq 0$ and that the baths include cutoff functions in frequencies, the real parts of the poles are negative (see Ref.\cite{CTPScalar}).

On the other hand, the EM retarded Green tensor is defined by the equations of motion obtained from the EM CTP action of Eq.(\ref{EqMotionAmuAlpha}) after imposing the temporal gauge.

In the inhomogeneous and anisotropic case, by solving Eq.(\ref{EqMotionEMRetGreenTensorLaplace}) for a given permittivity tensor defining boundaries and material bodies, we can obtain the Laplace transform of the EM retarded Green tensor and calculate the Casimir pressure through Eqs.(\ref{PICDoubleLaplace}), (\ref{PDOFDoubleLaplace}) and (\ref{PBathDoubleLaplace}). However, the solution in these cases is very complicated.

Therefore, we turn to the simpler case of considering the Lifshitz problem, where two different homogeneous and isotropic parallel half-spaces are separated by a vacuum gap of length $l$ along the $z$ direction. The origin of the coordinate system is defined in the middle of the gap. In this case, each contribution to the Casimir pressure has to be evaluated from Eqs.(\ref{PICDoubleLaplace}), (\ref{PDOFDoubleLaplaceUniformBodies}) and (\ref{PBathDoubleLaplaceUniformBodies}), where the last two equations are simplified versions of the general inhomogeneous and anisotropic case, but the first one remains unaltered. In this case, the Fresnel's ellipsoid turns out to be a sphere and the permittivity tensor results a function times the identity matrix, allowing any basis to describe each material body. Therefore, as we anticipated before, we omit the superscripts $[j]$ and replace $\mathbf{x}$ by the indices $L,R$ related to each material body. All in all, Eq.(\ref{EqMotionEMRetGreenTensorLaplace}) simplifies to:

\begin{equation}
\Big(\nabla\times\nabla\times+~s^{2}~\varepsilon(s,z)\Big)\overleftrightarrow{\mathcal{G}}_{\rm Ret}(\mathbf{x},\mathbf{x}',s)=-~\mathbb{I}~\delta(\mathbf{x}-\mathbf{x}'),
\label{EqMotionEMRetGreenTensorLaplaceLifshitz}
\end{equation}

\noindent where the refraction index only depends on $z$ and it is given by $$n^{2}(s,z)=\varepsilon(s,z)=1+\lambda_{0,L}^{2}~\Theta\left(-\frac{l}{2}-z\right)~G_{\rm Ret}^{L}(s)+$$ $$\lambda_{0,R}^{2}~\Theta\left(z-\frac{l}{2}\right)~G_{\rm Ret}^{R}(s).$$

It is remarkable that this equation and the one solved in Ref.\cite{Antezza} are basically the same, but this time the solution gives the Laplace transform of the retarded EM Green tensor. Two differences are in order: the equation is formally the same by replacing the Fourier variable in the solutions given in Ref.\cite{Antezza} by $is$; on the other hand, after the replacement, the r.h.s. of the Green function equation in Ref.\cite{Antezza} is the r.h.s. of Eq.(\ref{EqMotionEMRetGreenTensorLaplaceLifshitz}) times $4\pi s^{2}$, because it is compensating the differences in the definitions of the operator $\Theta^{sm}$ in the pressure expression of Eq.(\ref{PCasDoubleLaplace}) as it was commented at the end of Sect.\ref{EMCFP-TSP}. As the Laplace variable appears as a parameter in the equation, we can easily obtain the Laplace transform of the EM retarded Green tensor by dividing the one given in Ref.\cite{Antezza} by  $4\pi s^{2}$. Due to translational invariance in the parallel coordinates, the solution is given in terms of the Fourier transform in those coordinates:

\begin{equation}
\mathcal{G}_{\rm Ret}^{ij}(\mathbf{x},\mathbf{x}',s)=\int\frac{d\mathbf{Q}}{(2\pi)^{2}}~e^{i\mathbf{Q}\cdot\left(\mathbf{x}_{\parallel}-\mathbf{x}'_{\parallel}\right)}~\mathcal{G}_{\rm Ret}^{ij}(z,z',\mathbf{Q},s),
\label{LaplaceParallelFourierEMRetGreenTensor}
\end{equation}

\noindent where $\mathbf{Q}=(Q_{x},Q_{y},0)$, being $\check{x},\check{y}$ the parallel directions.

For a field point inside the gap ($-\frac{l}{2}<z<\frac{l}{2}$), the transform of the EM Green tensor depends on the value of the source point $z'$. For $z'<-\frac{l}{2}$ we have:

\begin{eqnarray}
&&\mathcal{G}_{\rm Ret}^{ij}(z,z',\mathbf{Q},s)=\frac{-1}{2q_{z}^{(1)}}\sum_{\mu=\rm TE,TM}\frac{t_{1}^{\mu}}{D_{\mu}}\Big[e_{\mu,i}[+]~e^{-q_{z}\left(z-\frac{l}{2}\right)}\nonumber\\
&&+~e_{\mu,i}[-]~r_{2}^{\mu}e^{q_{z}\left(z-\frac{l}{2}\right)}~e^{-2q_{z}l}\Big]e_{\mu,j}^{(1)}[+]~e^{q_{z}^{(1)}\left(z'-\frac{l}{2}\right)},
\label{LaplaceEMRetGreenTensorLeft}
\end{eqnarray}

\noindent and for $z'>\frac{l}{2}$:

\begin{eqnarray}
&&\mathcal{G}_{\rm Ret}^{ij}(z,z',\mathbf{Q},s)=\frac{-1}{2q_{z}^{(2)}}\sum_{\mu=\rm TE,TM}\frac{t_{2}^{\mu}}{D_{\mu}}\Big[e_{\mu,i}[-]~e^{q_{z}\left(z-\frac{l}{2}\right)}\nonumber\\
&&+~e_{\mu,i}[+]~r_{1}^{\mu}e^{-q_{z}\left(z-\frac{l}{2}\right)}\Big]e_{\mu,j}^{(2)}[-]~e^{-q_{z}l}~e^{-q_{z}^{(2)}\left(z'-\frac{3}{2}l\right)},
\label{LaplaceEMRetGreenTensorRight}
\end{eqnarray}

\noindent while for a point source inside the gap $-\frac{l}{2}<z'<\frac{l}{2}$ the tensor splits in bulk and scattered contributions ($\mathcal{G}_{\rm Ret}^{ij}=\mathcal{G}_{\rm Ret,Bu}^{ij}+\mathcal{G}_{\rm Ret,Sc}^{ij}$):

\begin{eqnarray}
&&\mathcal{G}_{\rm Ret,Bu}^{ij}(z,z',\mathbf{Q},s)=-\frac{\delta_{iz}~\delta_{jz}}{s^{2}}~\delta(z-z')\label{LaplaceEMRetGreenTensorBulk}\\
&&-\frac{1}{2q_{z}}\sum_{\mu=\rm TE,TM}\Big[e_{\mu,i}[+]~e_{\mu,j}[+]~e^{-q_{z}(z-z')}~\Theta(z-z')\nonumber\\
&&+~e_{\mu,i}[-]~e_{\mu,j}[-]~e^{q_{z}(z-z')}~\Theta(z'-z)\Big],\nonumber
\end{eqnarray}

\begin{eqnarray}
&&\mathcal{G}_{\rm Ret,Sc}^{ij}(z,z',\mathbf{Q},s)=-\frac{1}{2q_{z}}\sum_{\mu=\rm TE,TM}\frac{1}{D_{\mu}}\Big[e_{\mu,i}[+]~e_{\mu,j}[+]\nonumber\\
&&\times~r_{1}^{\mu}r_{2}^{\mu}~e^{-q_{z}(z-z'+2l)}+e_{\mu,i}[+]~e_{\mu,j}[-]~r_{1}^{\mu}e^{-q_{z}(z+z'-l)}\nonumber\\
&&+~e_{\mu,i}[-]~e_{\mu,j}[+]~r_{2}^{\mu}e^{q_{z}(z+z'-3l)}+e_{\mu,i}[-]~e_{\mu,j}[-]~r_{1}^{\mu}r_{2}^{\mu}\nonumber\\
&&\times~e^{q_{z}(z-z'-2l)}\Big].
\label{LaplaceEMRetGreenTensorScattered}
\end{eqnarray}

Here we are considering the notation given in Ref.\cite{Antezza} adapted to our case. Therefore, the EM (complex) wave vector in the medium $n$ is given by:

\begin{equation}
\mathbf{q}^{(n)}[\pm]=\mathbf{Q}\pm iq_{z}^{(n)}~\check{\mathbf{z}},
\end{equation}

\noindent with $n=L,R$ corresponding to each plate (while omitting the indices for the vacuum region) and where the sign $+$ ($-$) corresponds to an upward (downward) wave. The vector $\mathbf{Q}$ is always a real vector and appears as the projection of the wave vector $\mathbf{q}^{(n)}[\pm]$ on the interface, while the $z-$component is given by:

\begin{equation}
q_{z}^{(n)}=\sqrt{\varepsilon_{n}(s)~s^{2}+Q^{2}}.
\end{equation}

The wave vector $\mathbf{q}^{(n)}[\pm]$ lies in the plane of incidence defined by $\check{\mathbf{Q}}$ and $\check{\mathbf{z}}$. We can introduce the tranverse electric (TE) and magnetic (TM) polarization vectors:

\begin{equation}
\mathbf{e}_{\rm TE}^{(n)}[\pm]=\check{\mathbf{Q}}\times\check{\mathbf{z}},
\label{TEPolVector}
\end{equation}

\begin{equation}
\mathbf{e}_{\rm TM}^{(n)}[\pm]=\mathbf{e}_{\rm TE}^{(n)}[\pm]\times\check{\mathbf{q}}^{(n)}[\pm]=\frac{Q~\check{\mathbf{z}}\mp~iq_{z}^{(n)}~\check{\mathbf{Q}}}{\sqrt{\varepsilon_{n}(s)}~is}.
\label{TMPolVector}
\end{equation}

The reflection ($r$) and transmission ($t$) Fresnel coefficients for each single surface and components are given by:

\begin{equation}
r_{n}^{\rm TE}=\frac{q_{z}-q_{z}^{(n)}}{q_{z}+q_{z}^{(n)}}~~~~~,~~~~~r_{n}^{\rm TM}=\frac{\varepsilon_{n}~q_{z}-q_{z}^{(n)}}{\varepsilon_{n}~q_{z}+q_{z}^{(n)}},
\label{ReflectionCoeffInterface}
\end{equation}

\begin{equation}
t_{n}^{\rm TE}=\frac{2~q_{z}^{(n)}}{q_{z}+q_{z}^{(n)}}~~~~~,~~~~~t_{n}^{\rm TM}=\frac{2\sqrt{\varepsilon_{n}(s)}~q_{z}^{(n)}}{\varepsilon_{n}~q_{z}+q_{z}^{(n)}},
\label{TransmissionCoeffInterface}
\end{equation}

\noindent while the multiple reflections (which does not enter in the bulk part) are described by the denominator:

\begin{equation}
D_{\mu}=1-r_{1}^{\mu}~r_{2}^{\mu}~e^{-2q_{z}l}.
\label{Dmu}
\end{equation}

Beyond the apparent complicated expressions for the Laplace transform of the EM retarded Green tensor, its analytical properties can be analyzed relatively easy. Each expression presents several poles and branch cuts. The poles basically are the ones associated to $D_{\mu}$ and there is one pole located at the origin $s=0$.

The poles coming from $D_{\mu}$ should be calculated in order to obtain the time evolution of the system. However, causality implies that all these poles must lie on the left-half of the complex $s-$plane in order to give convergent contributions in the evolution. It may happen that some poles have vanishing real part. These poles would contribute to the long-time behaviour of the EM retarded Green tensor (and therefore to the pressure). However,  if we set $s=i~\text{Im}(s)$ (with $\text{Im}(s)\neq 0$) it can be easily shown that there is no poles of this type.

The pole at the origin appears explicitly in the first term of Eq.(\ref{LaplaceEMRetGreenTensorBulk}). However, in the second  term and in Eqs. (\ref{LaplaceEMRetGreenTensorLeft}), (\ref{LaplaceEMRetGreenTensorRight}) and (\ref{LaplaceEMRetGreenTensorScattered}), the pole is only associated to the terms with $\mu=\rm TM$, since in any region the $\rm TM$ polarization vectors provide the pole through its denominators, as is clear in Eq.(\ref{TMPolVector}).

Besides the poles, the Laplace transforms in each region present several branch cuts, which provide to the transient time evolution of the EM Green tensor. In the present case, by inspection, the branch cuts are given by the square roots $\sqrt{s^{2}+k_{\parallel}^{2}}$ and $\sqrt{\varepsilon_{n}(s)s^{2}+k_{\parallel}^{2}}$. The extra branch cut $\sqrt{\varepsilon_{n}(s)}$ that appears in the expression of the transmission coefficient for the $\rm TM$ terms in Eq.(\ref{TransmissionCoeffInterface}) that enter the Laplace transforms in Eqs.(\ref{LaplaceEMRetGreenTensorLeft}) and (\ref{LaplaceEMRetGreenTensorRight}) should not be considered, since it cancels with the same square root  contained in the denominator of the polarization vectors $\mathbf{e}_{\rm TM}^{(n)}(\pm)$ of Eq.(\ref{TMPolVector}), that also enter the same Laplace transforms.

Summarizing, the analytical properties of the Laplace transform of the EM retarded Green tensor are very simple, despite of its complicated expressions. These properties are important in order to compute the full time evolution of the problem and determining the long-time limit of the different physical quantities of interest. 

\subsection{The Steady Situation of the Lifshitz Problem}

Eqs.(\ref{LaplaceEMRetGreenTensorLeft})-(\ref{LaplaceEMRetGreenTensorScattered}) contain the Laplace-Fourier transforms for the EM retarded Green tensor for a field point inside the gap in the Lifshitz problem, from which we can calculate the corresponding Casimir pressure in the steady state.

For all the three contributions of Eqs.(\ref{PICDoubleLaplace}), (\ref{PDOFDoubleLaplaceUniformBodies}) and (\ref{PBathDoubleLaplaceUniformBodies}), in order to completely solve its time evolutions we have to calculate both inverse-Laplace transforms through the residue theorem and the analytical properties associated to branch cuts.

However, as our aim is to determine the long-time limit of the contributions, the work is simpler and  related to a subtle issue of the long-time regime. The `steadiness' of the physical quantities as the pressure or energy is related to the fact that they do not depend on time in the long-time regime. In other words, usually for the systems of interest, due to physical/intuitive reasons, the complete system is expected (or assumed) to be in a steady situation in the long-time limit. In many cases, this allows one to face and study the steady regime directly, without a rigorous (and unnecessary) derivation from a full dynamical problem. In the present case, we are in the opposite  situation. We have solved the full dynamical problem in terms of double inverse-Laplace transform, that should be calculated using the residue theorem. But we cannot do that completely due to the technical impossibility of locating the poles of the integrands in the complex planes analytically (we only know a few properties related to causality of the retarded Green tensors).

Nevertheless, to obtain the long-time limit of quantities given in terms of double inverse-Laplace transforms depending on $e^{(s_{1}+s_{2})(t_{1}-t_{i})}$ is relatively easier.

As the time evolution depends on the configuration of poles and branch cuts that the integrand presents, we have to distinguish which analytical properties contribute to the steady state of the quantities. As it is stressed in Ref.\cite{Jackson}, we can show that the typical branch cuts that appear in the Laplace transform of the EM retarded Green tensor do not contribute to the long-time regime. Then, the steady situation must be given by some of the poles, which always result in exponential temporal behaviours depending on the poles' values. For example, considering a typical Laplace transform (i.e., depending only on one Laplace variable), if a given pole have a non-zero real part, through causality we know that it must be a negative number. Therefore, the resulting behaviour associated to that pole will be an exponential decaying with time. On the other hand, if a given pole has vanishing real part (i. e., the pole is purely imaginary), the resulting behaviour will be oscillating in time with a frequency given by the imaginary part of the pole. As a last case, it turns out that if the pole is 0, the resulting behaviour will be of a time-independent constant.

However, in our case, the presence of $e^{(s_{1}+s_{2})(t_{1}-t_{i})}$ in the double inverse-Laplace transforms makes the poles associated to $s_{1}$ to combine with those associated to $s_{2}$ in order to determine the time evolution of the quantity. Thus, the case of a pole at $s=0$ is not the only way  we can obtain  time-independent behaviours in our quantities of interest. In other words, if we solve both Laplace integrals, each term of the full time evolution of the quantity will be a combination of two of the poles, one associated to the Laplace integral over $s_{1}$ (for example $\mathfrak{s}_{1}\in\mathbb{C}$) and the other one to $s_{2}$ (consider it at $s_2=\mathfrak{s}_{2}$). Then, if we combine two poles satisfying $\mathfrak{s}_{1}=-\mathfrak{s}_{2}$, there will be no exponential associated to that term in the time evolution, obtaining a time-independent term. In particular, is clear that the case of both poles equal to 0 satisfies the required condition, giving indeed a time-independent term. However, if we take a pole with negative real part, for example $\mathfrak{s}_{1}$ in such a way that $\text{Re}[\mathfrak{s}_{1}]<0$, the required condition to obtain a time-independent term would $\mathfrak{s}_{2}=-\mathfrak{s}_{1}$. As this implies $\text{Re}[\mathfrak{s}_{2}]>0$, which is forbidden by the causality property, then for all physical quantities builded from causal quantities, all the poles presenting negative real part cannot be combined in order to obtain a time-independent term.

As a last possibility, let us consider the poles with real part equal to 0. In these cases, considering a pole $i\mathfrak{s}_{1}$ with $\mathfrak{s}_{1}\in\mathbb{R}$, we can verify the required condition with a pole $i\mathfrak{s}_{2}=-i\mathfrak{s}_{1}$ in such a way that both oscillating evolutions cancel out in the exponential giving a time-independent term.

Following this train of thought, the time-independent terms will be so from the very beginning. The steady situation rises when the transient term vanish, i. e., when all the time-dependent terms resulting from poles combination and branch cuts go to 0. Thus, the relaxation time of the system will be defined by the last term that vanishes between all the transient terms. Moreover, the relaxation time will be equal to the smallest non-zero real part of the combination of poles or branch points in the three contributions.

All in all, in order to determine the steady state of each contribution it is necessary to study the combination of the poles present in both integrands over $s_{1}$ and $s_{2}$.

\subsubsection{The Long-Time Limit of the DOFs' Contribution}

We can start by considering one of the parts of the material's contribution to the Casimir pressure for the Lifshitz problem, which is associated to the DOFs' contribution given by the Eq.(\ref{PDOFDoubleLaplaceUniformBodies}) where, as we anticipated,  $n=L,R$. The quantity $$\lim_{\mathbf{x}_{2}\rightarrow\mathbf{x}_{1}}\Big[\Theta^{jk}(s_{1},s_{2})\int_{V_{n}}d\mathbf{x}\mathcal{G}_{\rm Ret}^{jb}(\mathbf{x}_{1},\mathbf{x},s_{1})\mathcal{G}_{\rm Ret}^{kb}(\mathbf{x}_{2},\mathbf{x},s_{2})\Big] ,$$ which is present in this contribution as well as in the baths' one can be simplified in the present case,  since in the last Section we have given the Laplace transform of the EM retarded Green tensor for the Lifshitz problem and its analytical properties.

Taking into account that for this problem the parallel coordinates can be Fourier-transformed as in Eq.(\ref{LaplaceParallelFourierEMRetGreenTensor}), the integrations over the parallel coordinates $\mathbf{x}_{\parallel}$ can be easily done, obtaining:

\begin{eqnarray}
&&\int_{V_{n}}d\mathbf{x}~\mathcal{G}_{\rm Ret}^{jb}(\mathbf{x}_{1},\mathbf{x},s_{1})~\mathcal{G}_{\rm Ret}^{kb}(\mathbf{x}_{2},\mathbf{x},s_{2})=\nonumber\\
&&=\int\frac{d\mathbf{Q}}{(2\pi)^{2}}~e^{i\mathbf{Q}\cdot\left(\mathbf{x}_{1\parallel}-\mathbf{x}_{2\parallel}\right)}\int_{V_{n}}dz~\mathcal{G}_{\rm Ret}^{jb}(z_{1},z,\mathbf{Q},s_{1})\nonumber\\
&&~~\times~\mathcal{G}_{\rm Ret}^{kb}(z_{2},z,-\mathbf{Q},s_{2}),
\end{eqnarray}

\noindent where in the r.h.s. the integration over $V_{n}$ implies the integration over each half-space. Then, we can write $$\int_{V_{n}}dz=(-1)^{n}\int_{(-1)^{n}\frac{l}{2}}^{(-1)^{n}\infty}dz, $$ where we associate $n=L$ ($n=R$) on the l.h.s. with $n=1$ ($n=2$) on the r.h.s. of this equality.

By considering Eqs.(\ref{LaplaceEMRetGreenTensorLeft}) and (\ref{LaplaceEMRetGreenTensorRight}), is clear that the remaining integration over $z$ involves exponential functions in the second argument, being easily calculated, obtaining:

\begin{eqnarray}
&&\int_{V_{n}}d\mathbf{x}~\mathcal{G}_{\rm Ret}^{jb}(\mathbf{x}_{1},\mathbf{x},s_{1})~\mathcal{G}_{\rm Ret}^{kb}(\mathbf{x}_{2},\mathbf{x},s_{2})=\label{IntegralEspacioGG}\\
&&\times\int\frac{d\mathbf{Q}}{(2\pi)^{2}}~e^{i\mathbf{Q}\cdot\left(\mathbf{x}_{1\parallel}-\mathbf{x}_{2\parallel}\right)}~\frac{1}{\left(q_{z}^{(n)}(s_{1},Q)+q_{z}^{(n)}(s_{2},Q)\right)}\nonumber\\
&&\times~\mathcal{G}_{\rm Ret}^{jb}(z_{1},(-1)^{n}l/2,\mathbf{Q},s_{1})~\mathcal{G}_{\rm Ret}^{kb}(z_{2},(-1)^{n}l/2,-\mathbf{Q},s_{2}).\nonumber
\end{eqnarray}

Therefore, from Eq.(\ref{PDOFDoubleLaplaceUniformBodies}), we finally obtain the full time evolution of the contribution to the Casimir pressure:

\begin{eqnarray}
&&P_{\rm DOFs}(x_{1},\beta_{P_{n}},l)=-\frac{1}{8\pi}\sum_{n=\rm L,R}\frac{\lambda_{0,n}^{2}M_{n}}{2\Omega_{n}}\coth\left[\frac{\beta_{P_{n}}\Omega_{n}}{2}\right]\nonumber\\
&&\times\int\frac{d\mathbf{Q}}{(2\pi)^{2}}\int_{\alpha_{1}-i\infty}^{\alpha_{1}+i\infty}\frac{ds_{1}}{2\pi i}\int_{\alpha_{2}-i\infty}^{\alpha_{2}+i\infty}\frac{ds_{2}}{2\pi i}~e^{(s_{1}+s_{2})(t_{1}-t_{i})}\label{PDOFDoubleLaplaceLifshitz}\\
&&\times\Big[\left(s_{1}^{2}~G_{\rm Ret,n}(s_{1})-1\right)\left(s_{2}^{2}~G_{\rm Ret,n}(s_{2})-1\right)+\Omega^{2}_{n}s_{1}s_{2}\nonumber\\
&&G_{\rm Ret,n}(s_{1})G_{\rm Ret,n}(s_{2})\Big]\lim_{\mathbf{x}_{2}\rightarrow\mathbf{x}_{1}}\Bigg[\Theta^{jk}(s_{1},s_{2})e^{i\mathbf{Q}\cdot\left(\mathbf{x}_{1\parallel}-\mathbf{x}_{2\parallel}\right)}\nonumber\\
&&\times~\frac{\mathcal{G}_{\rm Ret}^{jb}(z_{1},(-1)^{n}l/2,\mathbf{Q},s_{1})~\mathcal{G}_{\rm Ret}^{kb}(z_{2},(-1)^{n}l/2,-\mathbf{Q},s_{2})}{\left(q_{z}^{(n)}(s_{1},Q)+q_{z}^{(n)}(s_{2},Q)\right)}\Bigg],\nonumber
\end{eqnarray}

\noindent where we have stressed the fact that the contribution to the pressure depends on the plates separation $l$.

Now, in order to determine the long-time behaviour of this expression we have to analyze the integrand's analytical properties in terms of $s_{1}$ and $s_{2}$ simultaneously. For a given $n$, all the terms will not contribute necessarily. Taking in account the analytical properties of the retarded Green functions $G_{\rm Ret,n}$ and the EM retarded Green tensor $\overleftrightarrow{\mathcal{G}}_{\rm Ret}$ (which splits into $\rm TE$ and $\rm TM$ terms) commented in the last Section, by inspection it turns out that the only poles of the whole two-Laplace variables integrand resulting in a time-independent term are the ones located at the origin simultaneously for both variables. In other words, the poles located at the origin for both Laplace variables ($s_{1}=0=s_{2}$) associated to the products $\rm TM\times TM$ resulting from
\begin{equation} \label{GretGret}
\mathcal{G}_{\rm Ret}^{jb}(z_{1},(-1)^{n}l/2,\mathbf{Q},s_{1})~\mathcal{G}_{\rm Ret}^{kb}(z_{2},(-1)^{n}l/2,-\mathbf{Q},s_{2}).
\end{equation}
\noindent However this is not so straightforward because this product is also multiplied by $$[(s_{1}^{2}G_{\rm Ret,n}(s_{1}) -1)(s_{2}^{2}G_{\rm Ret,n}(s_{2})-1)+$$ $$\Omega^{2}_{n}s_{1}s_{2}~G_{\rm Ret,n}(s_{1})G_{\rm Ret,n}(s_{2})]$$ and $\Theta^{jk}(s_{1},s_{2})$,  containing linear factors $s_{1}$ and $s_{2}$ that can eventually prevent of $s_{1}=0$ or $s_{2}=0$ of being poles in a given term. At the end, we find poles of first and second order depending of which terms are considered.

As we mentioned before, the poles at the origin are related to the terms associated to the $\rm TM$ polarization vectors of Eq.(\ref{TMPolVector}). Therefore, from Eqs.(\ref{LaplaceEMRetGreenTensorLeft}) and (\ref{LaplaceEMRetGreenTensorRight}), we can see that the term $\rm TM\times TM$ of the product in Eq.(\ref{GretGret}) contains $1/s_{1}^{2}$ and $1/s_{2}^{2}$ as factors.

For example, terms associated to the combinations  $s_{1}^{2}$ $G_{\rm Ret,n}(s_{1})$ (or $s_{2}^{2}~G_{\rm Ret,n}(s_{2})$) will cancel out the denominators $1/s_{1}^{2}$ (or $1/s_{2}^{2}$) and therefore those terms will have no pole at $s_{1}=0$ ($s_{2}=0$). Those terms contribute only to the transient regime.

On the other hand, the term independent of $s_{1}$ and $s_{2}$ resulting from $$(s_{1}^{2}~G_{\rm Ret,n}(s_{1})-1)(s_{2}^{2}~G_{\rm Ret,n}(s_{2})-1)+$$ $$\Omega^{2}_{n}s_{1}s_{2}G_{\rm Ret,n}(s_{1})G_{\rm Ret,n}(s_{2})$$  will give poles of first order for both Laplace variables through it combination with the first term of $\Theta^{jk}(s_{1},s_{2})$ (associated to the electric field's contribution), and poles of second order for both variables in the combination with the second term of $\Theta^{jk}(s_{1},s_{2})$ (associated to the magnetic field's contribution).

Finally, the term $\Omega^{2}_{n}s_{1}s_{2}G_{\rm Ret,n}(s_{1})~G_{\rm Ret,n}(s_{2})$ will only present poles of first order when combined with $\Theta^{jk}(s_{1},s_{2})$ for the product with the second term (the combination with the first term will not have poles at the origin).

Therefore, the contribution to the pressure in Eq.(\ref{PDOFDoubleLaplaceLifshitz}) can be rewritten by grouping the terms according to the order of the pole at the origin in both variables:

\begin{eqnarray}
P_{\rm DOFs}(x_{1},\beta_{P_{n}},l)&=&(\text{Terms with second order poles at 0})\nonumber\\
&+&(\text{Terms with first order poles at 0})\nonumber\\
&+&(\text{Terms without poles at 0}).
\end{eqnarray}

In first place, it is clear that the last term has no contribution to the steady situation, taking part only in the transient stage.

In second place, it is easy to show that the second order poles at the origin result in terms that are directly zero or that diverge in the long-time limit ($t_{i}\rightarrow-\infty$). This is because, in the residue calculation of the second order poles, it is necessary to differentiate the integrand with respect to the Laplace variable in consideration (either $s_{1}$ or $s_{2}$). At this point, it is important to consider that the residue in these terms can be analyzed separately in each variable and that the integrands have the form of $e^{s(t-t_{i})}$ times a function which depend on $\sqrt{s^{2}+Q^{2}}$ and $\sqrt{\varepsilon_{n}(s)~s^{2}+Q^{2}}$. First, differentiating the exponential results in $(t_{1}-t_{i})e^{s(t_{1}-t_{i})}$ times the same function depending on $\sqrt{s^{2}+Q^{2}}$ and $\sqrt{\varepsilon_{n}(s)~s^{2}+Q^{2}}$. The evaluation of this on $s=0$ gives the mentioned term that diverge in the long-time limit. However, all these terms always have another term which cancels  them, giving no contribution. Secondly, differentiation the function depending on $\sqrt{s^{2}+Q^{2}}$ and $\sqrt{\varepsilon_{n}(s)~s^{2}+Q^{2}}$ gives terms that appear accompanied by $s/\sqrt{s^{2}+Q^{2}}$ or $[\varepsilon'_{n}(s)~s+2\varepsilon_{n}(s)]s/2\sqrt{\varepsilon_{n}(s)~s^{2}+Q^{2}}$, which evaluated on $s=0$ are the mentioned vanishing terms.

In conclusion, the second order poles at the origin do not contribute to the steady situation.

The remaining terms are the ones containing first order poles at the origin. These terms, in the temporal and spatial coordinates' domain, are associated to a time derivative of one of the Heaviside functions $\Theta(t-t_{i})$ which appear in the definition of the retarded Green functions or the EM retarded Green tensor due to its causal behaviour. Mathematically, deriving a Heaviside function gives a Dirac $\delta$-function. Therefore the expression is proportional to $\delta(0)$. However, this is caused by the sudden beginning of the interaction. Physically, this is an approximation to the fact that the interaction rises in a very short time. If instead, for example, we replace the Heaviside function by a smooth function going from 0 to 1 in finite time, approaching this value asymptotically, we would avoid this problem, keeping the convergent behaviour of the whole expression at the long-time limit. In summary, the first order poles at 0 are the result of deriving a Heaviside function representing the sudden switching-on of the interaction and hence they are not physical. Therefore, they will not have contribution to the long-time regime. This agrees with the limit for an analog contribution considered in Refs.\cite{CTPScalar,CTPGauge}, where the same contributions in the coordinates' domain are discarded by the same reasons.

Then, we have proved that for the contribution of the DOFs to the pressure, the long-time limit ($t_{i}\rightarrow-\infty$) is given by:

\begin{equation}
P_{\rm DOFs}(x_{1},\beta_{P_{n}},l)\longrightarrow P_{\rm DOFs}^{\infty}(\beta_{P_{n}},l)=0.
\label{PDOFSteadyLifshitz}
\end{equation}

\subsubsection{The Long-Time Limit of the Baths' Contribution}

Now we proceed to calculate the contribution resulting from the baths.

We recall the general expression of Eq.(\ref{PBathDoubleLaplaceUniformBodies}) found for the baths' contribution to the pressure in a scenario of homogeneous and isotropic bodies, considering as in the last case $n=\rm L,R$.

As in the DOFs' contribution, the pressure contains the expression $$\int_{V_{n}}d\mathbf{x}~\mathcal{G}_{\rm Ret}^{jb}(\mathbf{x}_{1},\mathbf{x},s_{1})~\mathcal{G}_{\rm Ret}^{kb}(\mathbf{x}_{2},\mathbf{x},s_{2}).$$

For the case of half-infinite parallel plates, Eq.(\ref{IntegralEspacioGG}) provides a simplification for the expression and then, the pressure reads:

\begin{eqnarray}
&&P_{\rm B}(x_{1},\beta_{n,\rm B},l)=\frac{-1}{8\pi}\sum_{n=\rm L,R}\lambda_{0,n}^{2}\int_{-\infty}^{+\infty}\frac{d\omega}{2\pi}\overline{N}_{n}(\omega)\int\frac{d\mathbf{Q}}{(2\pi)^{2}}\nonumber\\
&&\times\int_{\alpha_{1}-i\infty}^{\alpha_{1}+i\infty}\frac{ds_{1}}{2\pi i}\int_{\alpha_{2}-i\infty}^{\alpha_{2}+i\infty}\frac{ds_{2}}{2\pi i}~\frac{s_{1}s_{2}~e^{(s_{1}+s_{2})(t_{1}-t_{i})}}{(s_{1}+i\omega)(s_{2}-i\omega)}\nonumber\\
&&\times\frac{G_{\rm Ret,n}(s_{1})~G_{\rm Ret,n}(s_{2})}{\left(q_{z}^{(n)}(s_{1},Q)+q_{z}^{(n)}(s_{2},Q)\right)}\lim_{\mathbf{x}_{2}\rightarrow\mathbf{x}_{1}}\Big[\Theta^{jk}(s_{1},s_{2})\nonumber\\
&&\times~e^{i\mathbf{Q}\cdot\left(\mathbf{x}_{1\parallel}-\mathbf{x}_{2\parallel}\right)}~\mathcal{G}_{\rm Ret}^{jb}(z_{1},(-1)^{n}l/2,\mathbf{Q},s_{1})\nonumber\\
&&\times~\mathcal{G}_{\rm Ret}^{kb}(z_{2},(-1)^{n}l/2,-\mathbf{Q},s_{2})\Big].
\label{PBathDoubleLaplaceLifshitz}
\end{eqnarray}

Despite the branch cuts, in order to determine the long-time behaviour of this contribution we have to consider again combinations of poles in the Laplace variables $s_{1}$ and $s_{2}$ in such a way to obtain a time-independent term.

As a first step, as in the DOFs' contribution,  we have to analyze the pole at $s_{1}=0=s_{2}$ coming from the $\rm TM\times TM$ terms. However, due to the presence of a factor $s_{1}~s_{2}$ in the integrand, the only term of $\Theta^{jk}(s_{1},s_{2})$ that can have a contribution through this pole is the second one $\Lambda^{pm}~\epsilon^{prj}\epsilon^{mlk}~\partial_{r_{1}}\partial_{l_{2}}$ (associated to the magnetic fields), giving first order poles for both Laplace variables. On the other hand, the first term of $\Theta^{jk}(s_{1},s_{2})$ (associated to the electric fields) contains an extra factor $s_{1}~s_{2}$ which finally cancels the denominators in the terms $\rm TM\times TM$ which provide the pole at $s_{1}=0=s_{2}$, having no pole for the first term.

The calculation of the pole implies summing the indexes in the expressions. Calculating explicitly all the sums  and taking into account the coincidence limit, it is easy to see that the numerator also vanish for $s_{1}=0=s_{2}$. Then, when the denominator vanishes, the numerator vanishes too. Using L'Hopital rule one obtains a finite value, and therefore, $s_{1}=0=s_{2}$ it is not a pole.

Besides the point $s_{1}=0=s_{2}$, by inspection is clear that this contribution also presents another combination of poles on $s_{1}$ and $s_{2}$that produces a time-independent term. Due to the fact that the QBM noise kernel depends on the time difference (and then it can be written in terms of its Fourier transform), the denominators $(s_{1}+i\omega)$ and $(s_{2}-i\omega)$ appear in this contribution denoting the non-damped dynamics of the baths. Therefore, the pole at $s_{1}=-i\omega$ combined with the pole $s_{2}=i\omega$ gives a time-independent term.

All in all, the long-time contribution of the baths to the Casimir pressure can be written as:

\begin{eqnarray}
&&P_{\rm B}^{\infty}(\beta_{n,\rm B},l)=\frac{-1}{8\pi}\sum_{n=\rm L,R}\lambda_{0,n}^{2}\int_{-\infty}^{+\infty}\frac{d\omega}{2\pi}\omega^{2}\overline{N}_{n}(\omega)\int\frac{d\mathbf{Q}}{(2\pi)^{2}}\nonumber\\
&&\times\frac{G_{\rm Ret,n}(-i\omega)~G_{\rm Ret,n}(i\omega)}{\left(q_{z}^{(n)}(-i\omega,Q)+q_{z}^{(n)}(i\omega,Q)\right)}\lim_{\mathbf{x}_{2}\rightarrow\mathbf{x}_{1}}\Big[\Theta^{jk}(-i\omega,i\omega)\nonumber\\
&&\times~e^{i\mathbf{Q}\cdot\left(\mathbf{x}_{1\parallel}-\mathbf{x}_{2\parallel}\right)}~\mathcal{G}_{\rm Ret}^{jb}(z_{1},(-1)^{n}l/2,\mathbf{Q},-i\omega)\nonumber\\
&&\times~\mathcal{G}_{\rm Ret}^{kb}(z_{2},(-1)^{n}l/2,-\mathbf{Q},i\omega)\Big].
\label{PBathLongTimeLifshitz}
\end{eqnarray}

This is the contribution of the baths to the Casimir pressure in a general non-equilibrium context. An important connection with previous works can be established by employing a fluctuation-dissipation-type relation into the result. Besides that it is well-known that the fluctuation-dissipation theorem is valid for systems in thermal equilibrium, and the total system in this case is out of equilibrium, we can use a fluctuation-dissipation-type relation for the dissipation and noise kernel of the QBM ($D_{n}$ and $N_{n}$ respectively), i. e., the kernels generated by the baths in each point of space acting over each DOF. Therefore, we can write, for the Fourier transforms of the kernels, the fluctuation-dissipation relation:

\begin{equation}
\overline{N}_{n}(\omega)=\coth\left(\frac{\beta_{n,\rm B}}{2}~\omega\right)~\text{Im}\left[\overline{D}_{n}(\omega)\right],
\label{FluctuationDissipationRelationQBM}
\end{equation}

\noindent where $\overline{D}_{n}(\omega)$ is the Fourier transform of the QBM's dissipation kernel.

From the definition of the permittivity tensor in Eq. (\ref{PermittivityTensor}), it can be proved that for the present case, the Fourier transform of the permittivity function is given by $$\overline{\varepsilon}_{n}(\omega)=1+\lambda_{0,n}^{2}~\overline{G}_{\rm Ret,n}(\omega), $$ having that $\overline{G}_{\rm Ret,n}(-\omega)=\overline{G}_{\rm Ret,n}^{*}(\omega)$ for the reality of the QBM Green function. On the other hand, since the Laplace transform of the QBM $G_{\rm Ret,n}(s)$ of Eq.(\ref{QBMRetardedGreenFunction}) is assumed, by causality, to have poles with negative real parts, then it is verified that $G_{\rm Ret,n}(-i\omega)=\overline{G}_{\rm Ret}(\omega)$ and the same happens for the QBM dissipation kernel $D_{n}(-i\omega)=\overline{D}_{n}(\omega)$ (in fact, the connection between the Laplace and Fourier transforms applies for every causal function which is 0 for null value of its variable). Therefore, it is straightforward to prove that:

\begin{eqnarray}
\text{Im}\left[\overline{\varepsilon}_{n}(\omega)\right]&=&\text{Im}\left[\overline{D}_{n}(\omega)\right]~\left|\overline{G}_{\rm Ret,n}(\omega)\right|^{2}\label{RelationPermittivityGRetD}\\
&=&\text{Im}\left[\overline{D}_{n}(\omega)\right]~G_{\rm Ret,n}(-i\omega)~G_{\rm Ret,n}(i\omega).\nonumber
\end{eqnarray}

Then, introducing Eq.(\ref{FluctuationDissipationRelationQBM}) into Eq.(\ref{PBathLongTimeLifshitz}) and using Eq. (\ref{RelationPermittivityGRetD}), we obtain:

\begin{eqnarray}
&&P_{\rm B}^{\infty}(\beta_{n,\rm B},l)=-\frac{1}{8\pi}\sum_{n=\rm L,R}\int_{-\infty}^{+\infty}\frac{d\omega}{2\pi}~\omega^{2}\coth\left[\frac{\beta_{n,\rm B}\omega}{2}\right]\nonumber\\
&&\times~\text{Im}\left[\overline{\varepsilon}_{n}(\omega)\right]\int\frac{d\mathbf{Q}}{(2\pi)^{2}}\frac{1}{\left(q_{z}^{(n)}(-i\omega,Q)+q_{z}^{(n)}(i\omega,Q)\right)}\\
&&\times\lim_{\mathbf{x}_{2}\rightarrow\mathbf{x}_{1}}\Big[\Theta^{jk}(-i\omega,i\omega)~e^{i\mathbf{Q}\cdot\left(\mathbf{x}_{1\parallel}-\mathbf{x}_{2\parallel}\right)}\nonumber\\
&&\times\mathcal{G}_{\rm Ret}^{jb}(z_{1},(-1)^{n}l/2,\mathbf{Q},-i\omega)\mathcal{G}_{\rm Ret}^{kb}(z_{2},(-1)^{n}l/2,-\mathbf{Q},i\omega)\Big].\nonumber
\label{PBathLongTimeLifshitzCASIAntezza}
\end{eqnarray}

At this point, we  use Eq.(\ref{IntegralEspacioGG}) in order to go back to the spatial integral having $s_{1}=-i\omega$ and $s_{2}=i\omega$. Then, we can firstly employ the property associated to the reality of the EM retarded Green tensor in the time domain for the last factor $\mathcal{G}_{\rm Ret}^{ij}(\mathbf{x},\mathbf{x}',i\omega)=\mathcal{G}_{\rm Ret}^{*ij}(\mathbf{x},\mathbf{x}',-i\omega)$, followed by  the property of being a Feynman propagator to the last Laplace transform of the EM retarded Green tensor $\mathcal{G}_{\rm Ret}^{*ij}(\mathbf{x},\mathbf{x}',-i\omega)=\mathcal{G}_{\rm Ret}^{*ji}(\mathbf{x}',\mathbf{x},-i\omega)$. Finally,  for both Laplace transforms of the EM retarded Green tensor we use the connection between the Laplace and Fourier transform ensured by the causality behaviour to obtain:

\begin{eqnarray}
&&P_{\rm B}^{\infty}(\beta_{n,\rm B},l)=-\frac{1}{8\pi}\sum_{n=\rm L,R}\int_{-\infty}^{+\infty}\frac{d\omega}{2\pi}~\omega^{2}\coth\left[\frac{\beta_{n,\rm B}\omega}{2}\right]\nonumber\\
&&\times~\text{Im}\left[\overline{\varepsilon}_{n}(\omega)\right]\lim_{\mathbf{x}_{2}\rightarrow\mathbf{x}_{1}}\Big[\Theta^{jk}(\omega)\int_{V_{n}}d\mathbf{x}~\overline{\mathcal{G}}_{\rm Ret}^{jb}(\mathbf{x}_{1},\mathbf{x},\omega)\nonumber\\
&&\times~\overline{\mathcal{G}}_{\rm Ret}^{*bk}(\mathbf{x},\mathbf{x}_{2},\omega)\Big],
\label{PBathLongTimeLifshitzAntezza}
\end{eqnarray}

\noindent which is exactly the result obtained in Ref.\cite{Antezza} as the total pressure for the Lifshitz problem, where we have set $\Theta^{jk}(\omega)\equiv\Theta^{jk}(-i\omega,i\omega)$ from its definition.

All in all, we have proved that the baths' contribution in the long-time regime gives exactly the result found in Ref.\cite{Antezza} for the steady situation of the Lifshitz problem. However, following this procedure, it is worth noting that this result can be extended for the case of inhomogeneous and anisotropic materials since both purely imaginary poles are always present as it can be seen from the general expression of Eq.(\ref{PBathDoubleLaplace}) provided one can calculate the Laplace transform of the EM retarded Green tensor for a given problem, which is the main difficulty in most cases. Then, we can in general write:

\begin{eqnarray}
&&P_{\rm B}^{\infty}(\beta_{\mathbf{x},\rm B})=\frac{-1}{8\pi}\int_{-\infty}^{+\infty}\frac{d\omega}{2\pi}~\omega^{2}\lim_{\mathbf{x}_{2}\rightarrow\mathbf{x}_{1}}\Big[\Theta^{jk}(\omega)\int d\mathbf{x}~g(\mathbf{x})\nonumber\\
&&\times~\overline{\mathcal{G}}_{\rm Ret}^{jl}(\mathbf{x}_{1},\mathbf{x},\omega)\coth\left[\frac{\beta_{\mathbf{x},\rm B}}{2}~\omega\right]\text{Im}\left[\overline{\varepsilon}_{\mathbf{x}}^{[ll]}(\omega)\right]\nonumber\\
&&\times~\mathcal{G}_{\rm Ret}^{*lk}(\mathbf{x},\mathbf{x}_{2},\omega)\Big],
\label{PBathLongTimeGENERAL}
\end{eqnarray}

\noindent which is the full generalization of the steady result found in Ref.\cite{Antezza}.

\subsubsection{The Long-Time Limit of the Initial Conditions' Contribution}

After determining the contribution of the material to the Casimir pressure, the remaining contribution is the one associated to the EM field's initial conditions.

From Eq.(\ref{PICDoubleLaplace}), it is worth noting that unlike the other contributions this one does not simplify when considering homogeneous and isotropic bodies. In fact, it does not depend directly on the material properties and boundaries as the material's contributions do.  The dependence on the boundaries and the materials is encrypted by the EM retarded Green tensor but there is no more explicit dependence.

Nevertheless, for the particular case of the Lifshitz problem we can perform a  Fourier transform in the parallel coordinates through Eq.(\ref{LaplaceParallelFourierEMRetGreenTensor}) and then integrate them, obtaining:

\begin{eqnarray}
&&P_{IC}(x_{1},\beta_{EM})=-\frac{1}{8\pi}\int\frac{d\mathbf{k}}{2\omega_{\mathbf{k}}(2\pi)^{3}}\left[\delta^{bm}-\frac{k^{b}k^{m}}{\omega_{\mathbf{k}}^{2}}\right]\nonumber\\
&&\times\coth\left[\frac{\beta_{EM}\omega_{\mathbf{k}}}{2}\right]\int_{\alpha_{1}-i\infty}^{\alpha_{1}+i\infty}\frac{ds_{1}}{2\pi i}\int_{\alpha_{2}-i\infty}^{\alpha_{2}+i\infty}\frac{ds_{2}}{2\pi i}\left(s_{1}s_{2}+\omega_{\mathbf{k}}^{2}\right)\nonumber\\
&&\times~e^{(s_{1}+s_{2})(t_{1}-t_{i})}\lim_{\mathbf{x}_{2}\rightarrow\mathbf{x}_{1}}\Bigg[\Theta^{jk}(s_{1},s_{2})~e^{i\mathbf{k}_{\parallel}\cdot\left(\mathbf{x}_{1\parallel}-\mathbf{x}_{2\parallel}\right)}\nonumber\\
&&\times\int dz'~\mathcal{G}_{\rm Ret}^{jb}(z_{1},z',\mathbf{k}_{\parallel},s_{1})~e^{ik_{z}z'}\nonumber\\
&&\int dz''~\mathcal{G}_{\rm Ret}^{km}(z_{2},z'',-\mathbf{k}_{\parallel},s_{2})~e^{-ik_{z}z''}\Bigg],
\label{PICDoubleLaplaceParallelInt}
\end{eqnarray}

\noindent which is a simplification that can be done when the boundaries are parallel surfaces.

The integrals over $z'$ and $z''$  are over the whole axis $(-\infty,+\infty)$. Then, as the EM retarded Green tensor is given by Eqs.(\ref{LaplaceEMRetGreenTensorLeft})-(\ref{LaplaceEMRetGreenTensorScattered}) for each region, the integration separates into four integrals. For example, for the first integral we have:

\begin{eqnarray}
&&\int dz'~\mathcal{G}_{\rm Ret}^{jb}(z_{1},z',\mathbf{k}_{\parallel},s_{1})~e^{ik_{z}z'}=\nonumber\\
&&=\int_{-\infty}^{-l/2}dz'~\mathcal{G}_{\rm Ret}^{jb}(z_{1},z',\mathbf{k}_{\parallel},s_{1})~e^{ik_{z}z'}\nonumber\\
&&~~+\int_{-l/2}^{l/2}dz'~\mathcal{G}_{\rm Ret,Bu}^{jb}(z_{1},z',\mathbf{k}_{\parallel},s_{1})~e^{ik_{z}z'}\nonumber\\
&&~~+\int_{-l/2}^{l/2}dz'~\mathcal{G}_{\rm Ret,Sc}^{jb}(z_{1},z',\mathbf{k}_{\parallel},s_{1})~e^{ik_{z}z'}\nonumber\\
&&~~+\int_{l/2}^{+\infty}dz'~\mathcal{G}_{\rm Ret}^{jb}(z_{1},z',\mathbf{k}_{\parallel},s_{1})~e^{ik_{z}z'},
\end{eqnarray}

\noindent where the first and the last are the integrations having the source point in each plate, while the other two are the integrations
having the source point inside the gap.

As all the  transforms of the EM retarded Green tensors are given in term of exponentials, the integration is straightforward and can be written compactly as:

\begin{eqnarray}
&&\int dz'~\mathcal{G}_{\rm Ret}^{jb}(z_{1},z',\mathbf{k}_{\parallel},s_{1})~e^{ik_{z}z'}=-\frac{\delta_{j3}~\delta_{b3}}{s_{1}^{2}}~e^{ik_{z}z_{1}}\nonumber\\
&&+\sum_{n=1}^{2}\sum_{\mu}\Bigg\{\frac{(-1)}{2q_{z}\left[q_{z}+(-1)^{n-1}ik_{z}\right]}\Bigg(e_{\mu,j}\left[(-)^{n-1}\right]\nonumber\\
&&\times~e_{\mu,b}\left[(-)^{n-1}\right]e^{(-1)^{n}q_{z}z_{1}}\Big[e^{\left[(-1)^{n-1}q_{z}+ik_{z}\right]z_{1}}\nonumber\\
&&-e^{\left[-q_{z}+(-1)^{n}ik_{z}\right]\frac{l}{2}}\Big]+\frac{1}{D_{\mu}}\Big[e_{\mu,j}\left[(-)^{n-1}\right]r_{n}^{\mu}e^{q_{z}\left[(-1)^{n}z_{1}-2l\right]}\nonumber\\
&&+e_{\mu,j}\left[(-)^{n}\right]e^{q_{z}\left[(-1)^{n-1}z_{1}-l+(-1)^{n}2l\right]}\Big]e_{\mu,b}\left[(-)^{n-1}\right]r_{3-n}^{\mu}\nonumber\\
&&\times\Big[e^{\left[q_{z}+(-1)^{n-1}ik_{z}\right]\frac{l}{2}}-e^{-\left[q_{z}+(-1)^{n-1}ik_{z}\right]\frac{l}{2}}\Big]\Bigg)\nonumber\\
&&-\frac{e^{\left[(-1)^{n}ik_{z}-[(-1)^{n}+1]q_{z}\right]\frac{l}{2}}}{2q_{z}^{(n)}\left[q_{z}^{(n)}+(-1)^{n-1}ik_{z}\right]}\frac{t_{n}^{\mu}}{D_{\mu}}e^{(-1)^{n}q_{z}^{(n)}l}e_{\mu,b}^{(n)}\left[(-)^{n-1}\right]\nonumber\\
&&\times\Big[e_{\mu,j}\left[(-)^{n-1}\right]e^{(-1)^{n}q_{z}\left[z_{1}-\frac{l}{2}\right]}+e_{\mu,j}\left[(-)^{n}\right]r_{3-n}^{\mu}\nonumber\\
&&\times~e^{(-1)^{n-1}q_{z}\left[z_{1}-\frac{l}{2}\right]}~e^{\left[(-1)^{n}-1\right]q_{z}l}\Big]\Bigg\}.
\label{IntegralEMGRetZ}
\end{eqnarray}


It is clear that in order to obtain the result for the other integral $\int dz''~\mathcal{G}_{\rm Ret}^{km}(z_{2},z'',-\mathbf{k}_{\parallel},s_{2})~e^{-ik_{z}z''}$, we only have to make the replacements $z_{1}\rightarrow z_{2}$, $s_{1}\rightarrow s_{2}$, $j\rightarrow k$, $b\rightarrow m$, $\mathbf{k}_{\parallel}\rightarrow-\mathbf{k}_{\parallel}$ and $k_{z}\rightarrow-k_{z}$ in Eq.(\ref{IntegralEMGRetZ}).

Therefore, studying the analytical structure (poles and branch cuts) of the results of both integrals will give us the transient evolution and also the steady regime of the initial conditions' contribution to the Casimir pressure. The branch cuts present in the integrands can be considered by pairs (see Ref.\cite{Jackson}) and the regions in the complex plane where the integrands in each variable are multivaluated can be reduced to vertical intervals of finite length characterized by non-positive real parts. From Eq.(\ref{IntegralEMGRetZ}) we see that the branch cut interval with null real part corresponds to the straight line  $s=i~\rm{Im}(s)$ with $\rm{Im}(s)\in(-k_{\parallel},k_{\parallel})$, although it does not contribute to the steady state, it is important to know that it can be always considered as a finite-lenght interval in the complex plane.

Then, as for the material contributions, the key point to determine the long-time regime of the present contribution is to combine the poles in each variable in such a way that $e^{(s_{1}+s_{2})(t_{1}-t_{i})}$ gives no temporal dependence. Therefore, we must consider complementary poles.

At first sight, the pole at the origin in each variable (provided by the analytical structure of the retarded Green tensor) satisfies the requirement to give a steady term.

Now, to see which terms present this pole in both variables, we have to consider the combinations of the terms of $\Theta^{jk}(s_{1},s_{2})$ and $(s_{1}s_{2}+\omega_{\mathbf{k}}^{2})$. As it happened before, the terms $\rm{TM}\times\rm{TM}$ in the integrals product in Eq.(\ref{PICDoubleLaplaceParallelInt}) are the ones which give the poles at the origin in each Laplace variable. The different combinations finally give terms without poles at the origin and terms with poles of first and second order in both variables.

As for the material contributions, the terms without poles do not contribute to the residue calculation, while the terms with poles of second order at the origin result in terms which cancel out or directly vanish.

On the other hand, as it happened before, the poles of first order, in the temporal domain, correspond to derivatives of the Heaviside functions which gives Dirac $\delta$-functi- ons. This is associated to the sudden beginning of the interaction and then these terms give no additional physical information. They are only mathematical consequences of the description introduced for the switching-on of the interaction. Therefore, the poles of first order do not contribute either.

All in all, the pole at the origin for both variables does not contribute to the pressure associated to the initial conditions.

However, in this case, the poles at the origin are not the only ones that may contribute. The spatial integration over $z$ provides additional poles which are not present in the material's contribution. This poles provide the (3+1) EM generalization for the modified modes which are considered as an ansatz in Refs.\cite{LombiMazziRL,Dorota} and fully demonstrated in Ref.\cite{CTPScalar}, for the scalar case. In the present case, from Eq.(\ref{IntegralEMGRetZ}), it is clear that integration over $z$ results in additional denominators given by $q_{z}+(-1)^{n-1}ik_{z}$ and $q_{z}^{(n)}+(-1)^{n-1}ik_{z}$ depending if the integration is carried out over the vacuum gap or over the material plates respectively.

In fact, for the terms associated to the integrations over the material plates, the roots provided by the denominator $q_{z}^{(n)}+(-1)^{n-1}ik_{z}$, due to the presence of $\varepsilon_{n}(s)$ inside $q_{z}^{(n)}$, present negative real parts, in agreement with the causality properties and the fact that the field dissipates in such regions. For the cases where these roots are effectively poles of the integrand (which is not necessarily true as we will see below), This implies that the calculation of the residue will give exponential decays in time, resulting in terms with vanishing long-time limit.

On the other hand, it can be immediately seen that for the terms associated to integrations over the vacuum gap, the roots provided by $q_{z}+(-1)^{n-1}ik_{z}$ are $s=\pm i\omega_{\mathbf{k}}$ for $n=1$ and $k_{z}<0$ or for $n=2$ and $k_{z}>0$, having null real parts due to the free propagation of the field inside the gap. This shows that $s=\pm i\omega_{\mathbf{k}}$ could be poles in general, because the multivalued region for the integrand is given by the finite-lenght interval $s=i~\rm{Im}(s)$ with $\rm{Im}(s)\in(-k_{\parallel},k_{\parallel})$, and these poles are always located outside it since we always have $\omega_{\mathbf{k}}>k_{\parallel}$ for every $\mathbf{k}$. Moreover, as we anticipated before, these poles are associated to the mentioned modified modes for the EM field.

However, beyond the general case, for the present one we only have shown that $s=\pm i\omega_{\mathbf{k}}$ are roots of the denominator of certain terms for $n=1$ and $k_{z}<0$ or for $n=2$ and $k_{z}>0$. Moreover, for the Lifshitz problem, the given form of the Laplace transform of the retarded Green tensor makes that $s=\pm i\omega_{\mathbf{k}}$ are also roots of the numerator for the same cases of $n=1$ and $k_{z}<0$ or $n=2$ and $k_{z}>0$. Therefore, the limits of the integrands in each variable when the variable tends to $\pm i\omega_{\mathbf{k}}$ are indeterminate limits. Applying L'Hopital rule we can compute the limit giving a finite result. This means that, for the Lifshitz problem, despite $s=\pm i\omega_{\mathbf{k}}$ are roots of the denominator, they are not poles.  Therefore, there is no residue for this case and, moreover, there is no steady nor transient contribution for $s=\pm i\omega_{\mathbf{k}}$.

All in all, we have  proved that for the Lifshitz problem the initial conditions' contribution to the Casimir pressure vanish at the steady state ($t_{i}\rightarrow+\infty$):

\begin{equation}
P_{\rm IC}(x_{1},\beta_{\rm EM},l)\longrightarrow P_{\rm IC}^{\infty}(\beta_{\rm EM},l)=0.
\label{PICSteadyLifshitz}
\end{equation}

This result means that there are no modified modes for this problem that contribute to the long-time regime of the Lifshitz problem. Physically, the result expresses a relaxation process where the field dissipates during the transient stage until reaching a steady state in the long-time regime. Moreover, for the initial conditions' contribution, the process is given by the competition between the dissipation of the field on the material bodies present and the free fluctuation in the dissipationless regions (see next Section for further conclusions about these features). For the case of the Lifshitz problem, as the material regions are two half-spaces while the vacuum gap has finite length, dissipation wins over free fluctuation and there are no modified modes in the long-time regime. When interactions begin, the transient stage take place and the initial free field vacuum modes start to adapt to the presence of the material bodies (which has also a transient dynamics). During this process, the free fluctuations inside the gap attenuate due to the dissipation exerted by the material plates and no modified modes can raise, to finally settle at the steady state.


Finally, after calculating each contribution of Eqs.(\ref{PDOFSteadyLifshitz}), (\ref{PBathLongTimeLifshitzCASIAntezza}) and (\ref{PICSteadyLifshitz}), we can write the Casimir pressure in the steady state ($t_{i}\rightarrow-\infty$) as:


\begin{equation}
P_{\rm Cas}(x_{1})\longrightarrow P_{\rm Cas}^{\infty}=P_{\rm B}^{\infty}(\beta_{\rm B,n},l),
\end{equation}

\noindent i.e., the bath contribution at the long-time regime is the one that gives the total Casimir pressure for the composite system.


\section{Final Remarks and Outlook}

Having solved the Lifshitz problem from a well-defined initial condition problem, and achieved expressions for the three contributions to the total pressure for all times, we have shown that in the long-time limit $t_{i}\rightarrow-\infty$, the Casimir pressure is given only by the baths' contribution. We also proved that this corresponds to the only contribution considered in Ref.\cite{Antezza}, using an approach based on the macroscopic Maxwell equations and the source theory. Thus, we have presented a first principles calculation in the context of non-equilibrium quantum open systems that shows that the initial conditions are erased in the steady state.

Although this looks absolutely natural at first glance, the results in the previous works \cite{CTPScalar,CTPGauge} suggest that there should be important modifications for slabs of finite width. Indeed, for plates of finite width and for a $\delta$-plate in $1+1$ dimensions, the contributions of the initial conditions are present in the steady state \cite{CTPScalar} and agree with the one obtained from the modified modes of Refs.\cite{LombiMazziRL,Dorota}. Looking back to the situations analyzed in Refs.\cite{CTPScalar,CTPGauge}, and considering the results of the present paper, the whole picture becomes clear. It follows that a non-vanishing contribution of the initial conditions at the steady state is related to the existence of dissipationless regions of infinite size. The physical interpretation is the following. When there are regions of infinite size where the field dissipates, it happens that the free fluctuations in the dissipationless regions vanish in the long time limit because the damping generated by the dissipative regions overcome the free fluctuations. Therefore, the initial conditions' contribution to the pressure also vanishes. On the contrary, for  situations where the infinite regions are the ones without dissipation (while the dissipative regions are of finite size), the free field fluctuations are damped only in the dissipative regions, resulting in modified modes at the steady situation (as it happens in Ref.\cite{CTPScalar} for the appropriate cases).

In fact, these modes are associated to the poles $\pm i\omega_{\mathbf{k}}$ that result from the spatial integrations over the direction orthogonal to the boundaries. As we mentioned before, the contribution of the initial conditions can be matched with a contribution associated to homogeneous solutions in `steady' quantization schemes \cite{LombiMazziRL,Bechler,Dorota}, but here we are showing that the modified modes of that contribution are the result of the dynamical adaptation of the free field vacuum modes to the material boundaries. Therefore, on one hand, the modified modes become associated to the creation and annihilation operators of the initial free field. On the other hand, from the perspective of a `steady' canonical quantization scheme, the Hilbert space of the modified modes in the non-equilibrium situation is spanned by the same creation and annihilation operators that for a free field.

All in all, we have successfully proved the foundations of Lifshitz theory in non-equilibrium scenarios. To reach this goal, we have set up a first principles quantum dynamical problem of the EM field interacting with matter and subjected to uncorrelated initial conditions. We then solved the full dynamical problem and derived Lifshitz theory as the steady state by taking the long-time limit. The physical understanding of the contributions that enter in the full dynamical problem allowed us to gain powerful insights about the occurrence  of the steady state in different systems involving the interaction between quantum  fields and material bodies, that commonly appear in Casimir physics. It would be of high interest to generalize the results of the present work to situations of slabs of finite width, in order to quantify the relevance of the contribution of the initial conditions to the Casimir pressure in the long time limit. Work in this direction is in progress.

As a final comment, regarding the material controversy around the Drude and plasma models for the conduction electrons, we can say that the first principle microscopic approach could give fruitful results to bring light to the discussion. As we mentioned before, here we have proved the non-equilibrium Lifshitz foundations for insulator material plates, with permittivity functions that result from the bounded electrons modeled as polarization degrees of freedom. Including conduction electrons at the microscopic model requires an action for representing their dynamics. This may result in different analytical properties for the retarded Green functions, giving new contributions to the steady state depending on the material model considered and exposing in an explicit way the limitations of the Lifshitz formula. This is left as pending future work.


\begin{acknowledgement}
We would like to thank Matias Leoni and Alan Garbarz for useful comments and discussions about complex analysis. This work is supported by CONICET, UBA, and ANPCyT, Argentina.
\end{acknowledgement}

\end{document}